\documentclass[11pt]{article}
\usepackage{fullpage}
\usepackage[margin=1in]{geometry}

\usepackage{graphicx}
\usepackage{xcolor}
\usepackage{hyperref}
\usepackage{algorithm,algpseudocode}
\usepackage{algorithmicx}
\usepackage{multicol}
\usepackage{multirow}
\usepackage{hhline}
\usepackage{amsmath}
\usepackage{mathtools}
\usepackage[capitalize,noabbrev]{cleveref}
\usepackage{float}
\usepackage{array}
\usepackage{xspace}
\usepackage{subfigure}
\newcolumntype{L}[1]{>{\raggedright\let\newline\\\arraybackslash\hspace{0pt}}m{#1}}
\newcolumntype{C}[1]{>{\centering\let\newline\\\arraybackslash\hspace{0pt}}m{#1}}
\newcolumntype{R}[1]{>{\raggedleft\let\newline\\\arraybackslash\hspace{0pt}}m{#1}}

\RequirePackage[T1]{fontenc} 
\RequirePackage[tt=false, type1=true]{libertine} 
\RequirePackage[varqu]{zi4} 
\RequirePackage[libertine]{newtxmath}

\usepackage{nicefrac}
\usepackage{array}
\usepackage{makecell}

\newtheorem{ass}{\textbf{Assumption}}

\newtheorem{lem}{\textbf{Lemma}}
\newtheorem{thm}{\textbf{Theorem}}

\newcommand{\caln}{\mathcal{N}}

\newcommand{\cala}{\mathcal{A}}

\newcommand{\cali}{\mathcal{I}}

\newcommand{\calv}{\mathcal{V}}

\newcommand{\calc}{\mathcal{C}}

\newcommand{\bp}{\boldsymbol{p}}

\newcommand{\yinv}{y^{\texttt{inv}}\xspace}

\newcommand{\opt}{\texttt{OPT}\xspace}
\newcommand{\alg}{\texttt{ALG}\xspace}

\newcommand{\osp}{\texttt{OSP}\xspace}

\newcommand{\dpa}{\textsc{Dynamic}\xspace}
\newcommand{\spa}{\textsc{Static}\xspace}

\newcommand{\ogap}{\texttt{OAP}\xspace}
\newcommand{\oscc}{\texttt{OSCC}\xspace}

\newcommand{\ex}{\mathbb{E}}

\begin{document}

\title{Static Pricing for Online Selection Problem and its Variants}

\author{Bo~Sun\thanks{University of Waterloo. Email: {\tt bo.sun@uwaterloo.ca}.}
\and
Hossein~Nekouyan~Jazi\thanks{University of Alberta. Email: {\tt nekouyan@ualberta.ca}.}
\and
Xiaoqi~Tan\thanks{University of Alberta. Email: {\tt xiaoqi.tan@ualberta.ca}.}
\and
Raouf Boutaba\thanks{University of Waterloo. Email: {\tt rboutaba@uwaterloo.ca}.}
}

\begin{titlepage}
\maketitle

\thispagestyle{empty}

\begin{abstract}
This paper studies an online selection problem, where a seller seeks to sequentially sell multiple copies of an item to arriving buyers. We consider an adversarial setting, making no modeling assumptions about buyers' valuations for the items except acknowledging a finite support. In this paper, we focus on a class of static pricing algorithms that sample a price from a pre-determined distribution and sell items to buyers whose valuations exceed the sampled price. Such algorithms are of practical interests due to their advantageous properties, such as ease of implementation and non-discrimination over prices. 

Our work shows that the simple static pricing strategy can achieve strong guarantees comparable to the best known dynamic pricing algorithms. 
Particularly, we design the optimal static pricing algorithms for the adversarial online selection problem and its two important variants: the online assignment problem and the online selection with convex cost. 
The static pricing algorithms can even attain the optimal competitive ratios among all online algorithms for the online selection problem and the online assignment problem. 
To achieve these results, we propose an economics-based approach in the competitive analysis of static pricing algorithms, and develop a novel representative function-based approach to derive the lower bounds. 
We expect these approaches will be useful in related problems such as online matching.
\end{abstract}
\end{titlepage}

\maketitle
\section{Introduction}
In an online selection problem, a seller aims to sell multiple units of an item to a sequence of buyers that arrive online. Each buyer offers a price for purchasing one item upon arrival, and the seller immediately and irrevocably decides whether to accept or reject the offered price with the objective of maximizing the total selected prices from all arrivals. To devise online algorithms with performance guarantees, this problem has been studied under various assumptions about the problem instance. Among them, the secretary problem~\cite{gardner1970mathematical} assumes that prices arrive in a uniformly random order, and the prophet inequality~\cite{samuel1984comparison} assumes that prices are drawn from known distributions. Different from these assumptions based on statistical modeling, this paper focuses on an adversarial setting, where prices can take arbitrary values within a bounded support. Such a setting is of theoretical interest in classic online search problems~\cite{el2001optimal,lorenz2009optimal,jiang2021online} and the online knapsack problem~\cite{zhou2008budget,sun2022online}, and it can also find various applications in revenue management problems~\cite{ma2020algorithms,ball2009toward}.

We study the adversarial online selection problem through posted pricing mechanisms: the seller discloses a unit price for the item before each buyer's arrival. Acting as price takers, buyers decide to purchase one item if the utility (i.e., the buyer's valuation of the item minus the posted price) is non-negative. 
Our goal is to maximize the social welfare (i.e., the sum of the seller's revenue and the buyers' utility) in the posted pricing mechanism, which also aligns with maximizing the valuations of all selected buyers in the online selection problem. 

There has been a stream of literature that studies how to design posted prices to attain (near) optimal guarantees for adversarial online selection problems (e.g., \cite{zhang2017optimal,tan2020mechanism,ma2020algorithms}). Nearly all of these works rely on dynamic pricing strategies, which publish different prices for different buyers to make the best use of pricing power. However, dynamic pricing raises additional concerns in practice since it introduces \textit{price discrimination}, i.e., the same item is sold at different prices for different buyers who arrive at different times. This incentivizes buyers to strategically make purchase decisions. For example, in the airline ticket and hotel room booking problems, where dynamic pricing has been widely applied, each buyer may query the same flight ticket or hotel room multiple times, seeking a lower posted price. Price discrimination may be advantageous for the seller to maximize revenue in the short term. However, it is inherently unfair to buyers and diminishes the buyer experience, potentially leading to adverse effects in the long run.

In contrast, static pricing, which maintains a fixed price for all buyers, is considered a fairer pricing scheme~\cite{lechowicz2023time,sekar2017posted,balcan2008revenue}, and is easier to implement in practice. Moreover, static pricing has been shown to achieve guarantees comparable to optimal online algorithms under stochastic input assumptions (e.g., in the multi-unit prophet problem~\cite{chawla2023static}). In special cases of the adversarial setting, a (randomized) static pricing has even been shown to achieve optimal guarantees among all online algorithms in a single item setting~\cite{jiang2021online}.
In this paper, we continue this line of research and study static pricing for the \textit{multi-unit} online selection problem and its variants. Our key question is:
\begin{center}
\textit{How to design the best possible static pricing strategy, and can the static pricing achieve comparable performance to dynamic pricing in adversarial online selection problems? }    
\end{center}

In this work, we aim to answer this question by designing and analyzing static pricing algorithms for the multi-unit online selection problem (\osp) and its two variants: (i) the \textit{online assignment problem (\ogap)}, which involves selecting from multiple supply sources. Here, a seller maintains multiple items, each with multiple units for sale. Consequently, the decision involves not only selecting each buyer but also determining which item should be assigned to the selected buyer; and (ii) the \textit{online selection with convex cost (\oscc)}, which provides each unit of an item at a non-decreasing marginal production cost. Previous research has explored dynamic pricing algorithms for both \ogap~\cite{ma2020algorithms} and \oscc~\cite{tan2023threshold}. However, static pricing algorithms for \osp and its two variants have received comparatively less attention. This paper aims to fill this research gap.

\begin{table}[t]
\caption{State-of-the-art competitive ratio results for the online selection problem and its variants under posted price mechanisms. In all problems, the seller has $C$ identical units of each item, and the buyers' valuations are bounded within $[L,U]$. $\theta := U/L$ is the fluctuation ratio of the buyers' valuations. All lower bounds are derived in the large supply regime as $C \to\infty$. In \ogap, $\omega$ is the solution of $\frac{e^\omega}{e^\omega - 1} = \frac{\ln\theta}{1 -\omega}$, and $\frac{e^\omega}{e^\omega - 1} \to \frac{e}{e-1}$ as $\theta \to 1$. In \oscc, $\alpha_{\oscc}^{C}$ and $\alpha_{\oscc}^{\infty}$ are the competitive ratios defined in Theorem~$1$ in~\cite{tan2023threshold} and Theorem~$3$ in~\cite{tan2020mechanism}, respectively. $h(\cdot)$ is the convex conjugate function of the production cost function.}
\label{tab:sota}
\centering
\vspace{0.2cm}
\begin{tabular}{|c||c|c|cc|}
\hline
\multirow{2}{*}{Problem} & \multirow{2}{*}{Supply} & \multirow{2}{*}{\makecell{Lower bound}} & \multicolumn{2}{c|}{Upper bound} \\ \cline{4-5} 
 &  &  & \multicolumn{1}{c|}{Deterministic dynamic pricing} & Randomized static pricing \\ \hline\hline
\multirow{3}{*}{\osp} & $C\to 1$ & \multirow{3}{*}{$1+\ln\theta$} & \multicolumn{1}{c|}{$\theta$} & \makecell{$1+\ln\theta$ \cite{jiang2021online}} \\ \cline{2-2} \cline{4-5} 
 & $C \to \infty$ &  & \multicolumn{1}{c|}{\makecell{$1+\ln\theta$ \cite{zhou2008budget}}} & \multirow{2}{*}{\makecell{$1+\ln\theta$ \\
\textbf{Theorem~\ref{thm:osp-ub} (optimal)}}} \\ \cline{2-2} \cline{4-4} 
 & Finite $C$ &  & \multicolumn{1}{c|}{\textbf{Lemma~\ref{lem:ub-osp-dp} (tight)}} & \\ \hline\hline
\multirow{3}{*}{\ogap} & $C\to 1$ & \multirow{3}{*}{$\frac{e^\omega}{e^\omega - 1}$ }& \multicolumn{1}{c|}{$\theta$} & \makecell{$\frac{e^\omega}{e^\omega - 1}$ \cite{jiang2021online} \& \cite{ma2020algorithms}} \\ \cline{2-2} \cline{4-5} 
 & $C \to \infty$ &  & \multicolumn{1}{c|}{\makecell{$\frac{e^\omega}{e^\omega - 1}$ \cite{ma2020algorithms}}} & \multirow{2}{*}{\makecell{$\frac{e^\omega}{e^\omega - 1}$ \\
\textbf{Theorem~\ref{thm:oap-ub} (optimal)}}} \\ \cline{2-2} \cline{4-4}
 & Finite $C$ &  & \multicolumn{1}{c|}{\makecell{$\frac{e^\omega(1+C)(e^{1/C}-1)}{e^\omega - 1}$ \cite{ma2020algorithms}}} &  \\ \hline\hline
\multirow{2}{*}{\oscc} & $C\to\infty$ & \multirow{2}{*}{$\alpha_{\oscc}^{\infty}$} & \multicolumn{1}{c|}{\makecell{$\alpha_{\oscc}^{\infty}$~\cite{tan2020mechanism}}} & \multirow{2}{*}{\makecell{$1+\ln\frac{h(U)}{h(L)}$ \\
\textbf{Theorem~\ref{thm:oscc-ub} (tight)}}} \\ \cline{2-2} \cline{4-4}
 & Finite $C$ &  & \multicolumn{1}{c|}{\makecell{$\alpha_{\oscc}^{C}$~\cite{tan2023threshold}}} &  \\ \hline
\end{tabular}
\end{table}

\subsection{Our Contributions}
We make contributions along three fronts:

\smallskip 
\noindent\textbf{Optimal static pricing algorithms.} We design randomized static pricing algorithms for the multi-unit online selection problem (\osp), online assignment problem (\ogap), and online selection with convex cost (\oscc), and prove that all these algorithms attain the best possible competitive ratios among static pricing algorithms. Furthermore, the competitive ratios of \osp and \ogap also match the lower bounds of the problems, and thus are optimal among all online algorithms. Our main results are summarized in Table~\ref{tab:sota}. 
 
\smallskip 
\noindent\textbf{Insights.} In \osp and its two variants, static pricing algorithms can achieve competitive performance comparable to dynamic pricing algorithms. In particular, static pricing algorithms can attain better competitive ratios than deterministic dynamic pricing algorithms in \osp and \ogap since randomization helps eliminate the performance loss due to the discreteness of finite selection decisions. In \oscc, the convexity of the production cost strictly limits the power of randomization in static pricing, making static pricing algorithms perform worse than dynamic pricing algorithms. This can be seen as the cost a seller incurs in order to prevent price discrimination. 
Additionally, we observe that static pricing achieves optimal competitive performance for both social welfare and revenue maximization objectives. In contrast, dynamic pricing faces additional challenges in maximizing revenue, as high-valuation buyers may switch to lower posted prices, leading to revenue losses.

\smallskip 
\noindent\textbf{Techniques.} We propose new approaches to analyze the upper bound of the static pricing algorithm and the lower bound of the three studied problems. The competitive analysis of the static pricing algorithms is based on an economics-based approach under the posted pricing mechanism.
This approach generalizes the economics-based analysis of Ranking for online bipartite matching~\cite{eden2021economics} to additionally consider that (i) each item has multiple copies, and (ii) valuations are both item and buyer dependent.
Our lower bound proofs are based on a novel representative function-based approach specifically proposed for online selection problems. Different from the classic method based on Yao's Minmax principle, our approach is not only general and applicable to \osp and its variants, but also guides the design of the distribution of the static price.



\subsection{Related Work}
The online selection problem has been extensively studied in a large body of literature, with researchers exploring various assumptions about problem inputs. For instance, the problem has been formulated as the secretary problem under the random-order model~\cite{gardner1970mathematical}, and as the prophet inequality problem under a stochastic input model~\cite{samuel1984comparison}. In contrast to these assumptions based on statistical models, our focus lies on the adversarial setting, where arrivals' valuations can be arbitrary within a bounded support. We will now delve into reviewing the related work concerning adversarial online selection problems.

\smallskip 
\noindent\textbf{Online selection problem (\osp)} 
The adversarial \osp has been extensively studied in the large supply regime (i.e., $C \to\infty$). \cite{zhou2008budget} first introduced an online knapsack problem with infinitesimal item weights, which includes our \osp in the large supply regime as a special case. They proposed a threshold-based algorithm that has been proven to achieve the optimal competitive ratio among all online algorithms. Subsequent works have shown that the threshold-based algorithm can be interpreted as a deterministic dynamic pricing algorithm~\cite{zhang2017optimal}, and have extended the algorithms and results from a single knapsack problem to multiple multi-dimensional knapsack problems~\cite{sun2022online}. However, all these works focus on dynamic pricing based on the infinitesimal assumption on the item size.
In contrast to these works, a recent study by \cite{jiang2021online} designs an optimal randomized static pricing algorithm for \osp in the unit supply regime (i.e., $C = 1$). This paper continues this line of work and aims to design static pricing algorithms for general $C$ settings. 

\smallskip 
\noindent\textbf{Online assignment problem (\ogap)} 
\ogap extends the \osp by considering multiple items, each with multiple identical copies. This model was initially studied in the online edge-weighted matching literature within the context of the Ad assignment problem~\cite{feldman2009online}. Achieving bounded competitive results in the Ad assignment problem relies on the free disposal assumption, which may not hold true in the general case. \cite{zhou2008budget} and \cite{sun2022online} studied this problem under the bounded valuation assumption (which is also assumed in this paper) in the large supply regime, framing it as online multiple knapsack problems. They designed dynamic pricing algorithms capable of achieving an order-optimal competitive ratio. \cite{ma2020algorithms} further improved the pricing function design, achieving state-of-the-art results that attain an exact optimal competitive ratio in the large supply regime and asymptotic optimal competitive ratio for the finite supply case.

In terms of static pricing, both \cite{ma2020algorithms} and \cite{jiang2021online} proposed optimal static pricing algorithms (framed as Ranking-like algorithms) for \ogap in the unit supply regime. However, in the finite supply regime, \cite{ma2020algorithms} suggested transforming multiple units of the same item into multiple single-unit items, followed by the utilization of a Ranking-like algorithm to solve the transformed problem. This transformation not only increases the number of static prices, with one price for each unit of an item, but also reintroduces price discrimination, as the prices of different units from the same item are sampled from the same distribution and may be provided to buyers differently.
In contrast to~\cite{ma2020algorithms}, our proposed algorithm sets a fixed price for each item, which avoids the concern of price discrimination. Additionally, we adopt an economics-based approach to analyze the algorithm, as opposed to the classic online primal-dual approach.

\smallskip 
\noindent\textbf{Online selection with convex cost (\oscc)}
\oscc is another extension of \osp that accounts for a non-linear supply cost. This setting was first studied by \cite{blum2011welfare} and \cite{huang2019welfare} in the general context of combinatorial auctions in the large supply regime. In the \oscc setting, \cite{tan2020mechanism} first designed an optimal dynamic pricing algorithm, which was later extended to a finite supply setting, achieving a tight competitive ratio (i.e., the optimal competitive ratio among all deterministic dynamic pricing algorithms). However, to the best of our knowledge, no static pricing algorithms with guaranteed performance are known for this problem.





\section{Online Selection Problem}

We study the online selection problem (\osp) through the lens of posted pricing.
In this problem, a seller aims to sell $C$ identical units of an item to $N$ buyers that arrive one at each time.
Upon the arrival of buyer $n$, the seller publishes the price $p_n$. Buyer $n$ has a private valuation $v_n$ on the item, and decides to take the item at a price $p_n$ if the valuation is higher than the price ($v_n \ge p_n$), and otherwise the buyer leaves without purchasing.
Let $x_n\in\{0,1\}$ denote the decision of the buyer $n$. Each buyer $n$ can obtain a utility $u_n = (v_n - p_n) x_n$, and the seller can collect a total revenue of $r = \sum_{n\in[N]} x_n p_n$.
The objective of the problem is to maximize the social welfare of the seller and all buyers, which is equivalent to maximizing $r + \sum_{n\in[N]} u_n = \sum_{n\in[N]}v_n x_n$. 

Let $I :=\{v_n\}_{n\in[N]}$ denote an instance of \osp that contains a sequence of the buyers' valuations. Given $I$, the offline optimal algorithm can determine the posted prices such that $\min\{C,N\}$ items are sold to the buyers with the maximum valuations. Let $\opt(I)$ denote the social welfare of the offline optimal algorithm and it can be obtained by solving the following problem
\begin{align}
    \max_{x_n} \quad\sum\nolimits_{n\in[N]} v_n x_n,\quad {\rm s.t.} \quad\sum\nolimits_{n\in[N]} x_n \le C,\quad  x_n \in \{0,1\}, \forall n\in[N].
\end{align}
In the online setting, the posted price for each buyer $n$ must be determined without knowing the information of future buyers $\{v_i\}_{i> n}$ and the total number of buyers $N$. Let $A$ denote an online  algorithm that determines the posted price $p_n := A(\{x_i\}_{i< n})$ just based on observed decisions of the previous $n-1$ buyers.
Let $\alg(A,I)$ denote the social welfare obtained by algorithm $A$ under the instance $I$. 
The performance of the online algorithm is quantified by the competitive ratio $\alpha(A) = \max_{I} \frac{\opt(I)}{\ex [\alg(A,I)]}$,
which is the worst-case ratio of the social welfare between offline optimal algorithm and the online algorithm, and the expectation is taken over the algorithm's randomness. Then our goal is to design the pricing algorithm $A$ that can minimize the competitive ratio.

Without additional information, there is no online algorithm that can achieve bounded competitive ratio for \osp. In this paper, we assume the valuations of buyers are bounded. 
\begin{ass}\label{ass:bounded-value}
Buyers' valuations in \emph{\osp} are bounded, i.e., $v_{n} \in [L, U], \forall n\in[N]$.
\end{ass}
The \osp under above assumption can be framed as an online optimization problem with predictions, where the valuations of buyers are predicted to fall within the interval $[L,U]$ with high probability ~\cite{jiang2021online}. The competitive ratios of online algorithms for \osp depend on the fluctuation ratio $\theta := U/L$ of the valuations, which indicate the intrinsic uncertainty of the problem.

\subsection{Dynamic Pricing for Online Selection Problem}
\label{sec:ppa-osp}

We first present a dynamic pricing algorithm (\dpa)
described in Algorithm~\ref{alg:ppa-osp}.
This algorithm takes as input a deterministic pricing function $\phi(z): [C] \to [L,U]$, where $\phi(z)$ is the posted price for the $z$-th unit of the item to be sold. In the limiting case when $C\to \infty$, prior work~\cite{zhou2008budget} has designed a pricing function  
\begin{align}
\label{eq:threshold1}
\phi_{\osp}^{\infty}(z) = 
\begin{cases}
L & z\in[0,C/\alpha_{\osp}^{\infty})\\
L \exp(\alpha_{\osp}^{\infty} z/C - 1), & z\in [C/\alpha_{\osp}^{\infty}, C]
\end{cases}, \ \text{with}\ \alpha_{\osp}^{\infty} = 1 + \ln\theta.
\end{align}
In this large supply regime (i.e., $C\to\infty$), the change in the number of available items is infinitesimal compared to the total number of items, and thus the pricing function is approximately a continuous function. Based on this property, \cite{zhou2008budget} has shown that the \dpa with $\phi_{\osp}^{\infty}$ as the pricing function can achieve the optimal competitive algorithm among all online algorithms. However, in the regime of finite supply $C$, the competitive analysis of dynamic pricing in \cite{zhou2008budget} does not hold. To the best of our knowledge, there is still no dynamic pricing algorithm in this finite $C$ regime. Therefore, we first design a pricing function $\phi^{C}_{\osp}$ for the \osp with $C$ items.

\begin{algorithm}[t]
\caption{Deterministic Dynamic Pricing for Online Selection Problem ($\dpa(\phi)$)}
\label{alg:ppa-osp}
\begin{algorithmic}[1]
\State \textbf{input:} pricing function $\phi(\cdot)$; 
\State \textbf{initiate:} $z = 1$;
\For{$n=1,\dots,N$} \Comment{buyer $n$ arrives with a private valuation $v_n$}
\State post a price $p_n = \phi(z)$; \Comment{publish a price using pricing function $\phi(\cdot)$}
\If{$v_n \ge p_n$ and $z\le C$} \label{alg:line-capacity}
\State set $x_n = 1$ and $z = z +1$; \Comment{buyer $n$ accepts the posted price}
\Else
\State set $x_n = 0$. \Comment{buyer $n$ rejects the posted price}
\EndIf
\EndFor
\end{algorithmic}
\end{algorithm}

\begin{lem}\label{lem:ub-osp-dp}
A deterministic dynamic pricing algorithm $\dpa(\phi^{C}_{\osp})$ is $\alpha^{C}_{\osp}$-competitive for the online selection problem when the pricing function is given by 
\begin{align}
\label{eq:threshold-osp}
\phi^{C}_{\osp}(z) = 
\begin{cases}
 L, &\text{if}\ z = 1,\dots, \lceil C/\alpha^{C}_{\osp} \rceil := \gamma, \\ \frac{\gamma L \alpha^{C}_{\osp}}{C} \left(1 + \frac{\alpha^{C}_{\osp}}{C} \right)^{z-\gamma-1}, & \text{if}\ z = \gamma+1,\dots, C, 
\end{cases}
\end{align}
where $\alpha^{C}_{\osp}$ is the solution of $\left(1 + \frac{\alpha}{C} \right)^{C-\gamma} = \frac{C \theta}{\gamma \alpha}$. 
Further, $\alpha^{C}_{\osp}$ is the minimum competitive ratio that can be attained by deterministic online algorithms. 
\end{lem}
The detailed proof of Lemma~\ref{lem:ub-osp-dp} is provided in Appendix~\ref{app:lem-osp-dp}. Note that as $C\to\infty$, $\alpha^{C}_{\osp} \to \alpha^{\infty}_{\osp} = 1 + \ln\theta$, and $\alpha^{C}_{\osp}$ is strictly larger than $1+\ln\theta$. The gap between $\alpha^{C}_{\osp}$ and $\alpha^{\infty}_{\osp}$ arises because only a finite number of pricing decisions can be made when only $C$ units of the item are available for sale. In the following section, we demonstrate that such a gap can be eliminated through randomization. Furthermore, we establish that it is possible to achieve a competitive ratio of $\alpha^{\infty}_{\osp}$ using only randomized static pricing. 


\subsection{Static Pricing for Online Selection Problem}
\label{sec:osp-spa}

\begin{algorithm}[t]
\caption{Static Pricing for Online Selection Problem ($\spa(\psi)$)}
\label{alg:osp-spa}
\begin{algorithmic}[1]
\State \textbf{input:} inverse cumulative density function $\psi(\cdot)$; 
\State sample $x$ uniformly within $[0,1]$;
\State \textbf{initiate:} $z = 1$;
\For{buyer $n = 1,\dots, N$} \Comment{buyer $n$ arrives with a private valuation $v_n$}
\State post a price $p = \psi(x)$; \Comment{publish a static price sampled based on $\psi(\cdot)$}
\If{$v_n \ge p$ and $z\le C$}
\State set $x_n = 1$ and $z = z +1$; \Comment{buyer $n$ accepts the posted price}
\Else
\State set $x_n = 0$.  \Comment{buyer $n$ rejects the posted price}
\EndIf
\EndFor
\end{algorithmic}
\end{algorithm}

We next consider static pricing in the realm of posted price mechanisms. First, we assert that a deterministic static pricing can only achieve a competitive ratio of $\theta$.
This is evident because any static price set above $L$ results in unbounded competitive ratios, as all buyers may value the item slightly below the static price, leading to no sales. Therefore, the static price must be set at $L$. Consequently, we can construct an instance where $C$ buyers have valuations of $L$ followed by $C$ buyers with valuations of $U$. The worst-case ratio for this static pricing algorithm is $U/L = \theta$.
Therefore, our focus shifts to the randomized static pricing algorithm, where we can demonstrate that randomization can improve the competitive ratio to $1+\ln\theta$, the best achievable competitive ratio for any online algorithm (see the lower bound result in Section~\ref{sec:osp-lb}).

We present a randomized static pricing algorithm (\spa) in Algorithm~\ref{alg:osp-spa}.
This algorithm draws a uniform random variable $x$ within $[0,1]$ in the beginning, and then sets $p = \psi(x)$ as the fixed posted price for all buyers, where $\psi$ is the inverse cumulative density function (CDF) of the static price. 
We denote the static pricing algorithm by $\spa(\psi)$.

\begin{thm}
\label{thm:osp-ub}
A randomized static pricing algorithm $\spa(\psi)$ can achieve a competitive ratio $\alpha = \alpha_{\osp}^{\infty} = 1 + \ln\theta$ if the inverse CDF of the static price is given by
\begin{align}
\label{eq:threshold}
\psi(x) = 
\begin{cases}
L & x\in[0,1/\alpha)\\
L \exp(\alpha x - 1), & x\in [1/\alpha, 1]
\end{cases}.
\end{align}
\end{thm}
Based on Theorem~\ref{thm:osp-ub}, $\spa(\psi)$ not only avoids the price discrimination of dynamic pricing, but also attains a competitive ratio better than that of the best possible deterministic dynamic pricing strategy.
In the following, we prove Theorem~\ref{thm:osp-ub} by employing an economics-based analysis, which proves useful for studying the generalization of \osp in subsequent sections.

\paragraph{Proof of Theorem~\ref{thm:osp-ub}.}
$\spa(\psi)$ sets a price $p = \psi(x)$ by sampling from the inverse CDF $\psi$. Upon the arrival of buyer $n$, if her valuation $v_n \ge p$ and there are items available, the buyer purchases the item at the price $p$. We omit the subscript of $\psi$ in the proof when the context is clear. 
Given an arrival instance $I = \{v_1,\dots,v_N\}$, let $X_n \in \{0,1\}$ denote the online decision of $\spa(\psi)$.
Then each buyer $n$ obtains a utility $u_n = X_n(v_n - p)$, and the seller collects a total revenue $r = \sum_{n\in[N]} X_n p := \rho(x) \psi(x)$, where $\rho(x):= \sum\nolimits_{n\in[N]} X_n$ denotes the total number of sold items by \spa when the static price is given by $p = \psi(x)$.
The expected social welfare attained by $\spa(\psi)$ under instance $I$ can be described by
\begin{align}\label{eq:osp-ub-eq1}
\mathbb{E}[\alg(p,I)] &= 
\ex [\sum\nolimits_{n\in[N]} u_n + r] \ge \ex [r] = \int_{0}^1 \rho(x) \psi(x) dx,
\end{align}
where the inequality is due to the fact that $u_n \ge 0, \forall n\in[N]$.

To establish the connection between offline algorithm and online algorithm for a given instance $I$, let $(v_1^*, v_2^*, \dots, v_{N^*}^*)$ denote a sequence of buyers' valuations in $I$ that are selected by offline algorithm and arranged in a non-decreasing order. Thus, this sequence contains the 
$N^*$-maximum valuations from all buyers in $I$, where $N^* = \min\{C,N\}$, and $v_1^*$ is the minimum valuation selected. Then the optimal social welfare is $\opt(I) = \sum_{i\in [N^*]} v_i^*$.
Define a sequence of non-decreasing thresholds $(x_0, x_1,\dots, x_{N^*})$ such that $x_0 = 0$ and $\psi(x_i) = v_i^*, i\in[N^*]$.
We can lower bound $\rho(x)$ by 
\begin{align}\label{eq:osp-ub-eq2}
    \rho(x) = \sum\nolimits_{n\in[N]}X_{n} \ge \sum\nolimits_{i\in [N^*]} \mathbb{I}\{x \le x_i\}, \forall x\in[0,1],
\end{align}
where $\mathbb{I}\{x \le x_i\}$ is the indicator function, which equals $1$ when $x \le x_i$ and $0$ otherwise.
To see the above equation \eqref{eq:osp-ub-eq2}, note that if \spa with price $\psi(x)$ does not sell all $C$ items after the execution of the instance $I$, then all buyers with valuations greater than $\psi(x)$ make the purchase, and thus the above inequality holds. On the other hand, if \spa sells all items, then $\rho(x) = C \ge N^* \ge \sum\nolimits_{i\in [N^*]} \mathbb{I}\{x \le x_i\}$ since the offline algorithm can at most select $N^*$ items.

Based on equations~\eqref{eq:osp-ub-eq1} and~\eqref{eq:osp-ub-eq2}, we lower bound the expected social welfare of $\spa(\psi)$ by
\begin{subequations}
\begin{align}
\label{eq:ineq1}
\mathbb{E}[\alg(p,I)] &\ge  \int_{0}^{x_1} N^* \psi(x) dx  + \sum_{j=2}^{N^*}\int_{x_{j-1}}^{x_{j}} \left(\sum_{i = j}^{N^*} 1\right) \psi(x) dx,\\
\label{eq:ineq2}
& = N^*\int_{0}^{x_1}  \psi(x) dx + \sum_{i=2}^{N^*} \int_{x_{1}}^{x_{i}} \psi(x)dx \\
\label{eq:ineq3}
& \ge  N^*\frac{\psi(x_1)}{\alpha} + \sum\nolimits_{i=2}^{N^*} \frac{\psi(x_i) - \psi(x_1)}{\alpha}\\
&= \frac{1}{\alpha} \sum\nolimits_{i\in[N^*]} v_i^* = \frac{1}{\alpha} \opt(I),
\end{align}    
\end{subequations}
where the inequality~\eqref{eq:ineq1} results from equation~\eqref{eq:osp-ub-eq2}, equality~\eqref{eq:ineq2} is obtained by exchange of summation, and the inequality~\eqref{eq:ineq3} holds since the proposed inverse CDF $\psi$ in equation~\eqref{eq:threshold} gives
\begin{subequations}
\begin{align*}
    \int_{0}^{x_1} \psi(x)dx \ge \frac{\psi(x_1)}{\alpha} = \frac{v_1^*}{\alpha}, \quad\text{and}\quad   \int_{x_1}^{x_i} \psi(x) dx \ge \frac{1}{\alpha} \int_{x_1}^{x_i} \psi'(x) dx = \frac{v_i^* - v_1^*}{\alpha}, \quad i = 2,\dots, N^*. 
\end{align*}    
\end{subequations}
This completes the proof of Theorem~\ref{thm:osp-ub}. \hfill{$\square$}\smallskip

The missing piece of the static pricing algorithm is how to design the inverse CDF function $\psi$ to attain the optimal competitive ratio and why the function $\psi$ happens to be in the same form as the dynamic pricing function $\phi$ in the limiting case of dynamic pricing algorithms {(i.e., comparing equations \eqref{eq:threshold1} and \eqref{eq:threshold})}, although $\psi$ and $\phi$ have different meanings. We partially answer these two questions in the lower bound proof of \osp.




\subsection{Lower Bound of Online Selection Problem}
\label{sec:osp-lb}

We prove a lower bound result for the online selection problem using a representative function approach.
Although the lower bound can also be proved using the classic approach based on Yao's minmax principle (based on a slight modification of the lower bound proof in~\cite{zhou2008budget}), we emphasize that the representative function-based approach can guide the design of the static pricing algorithm. Moreover, it is specifically designed for the \osp and can be extended to prove lower bound results for more general settings in subsequent sections.
\begin{lem}
\label{lem:osp-lb}
No online algorithm, including deterministic and randomized algorithms, for the online selection problem can achieve a competitive ratio smaller than $1 + \ln\theta$.
\end{lem}

The high-level idea for proving the lower bound is to \textit{first} construct a family of hard instances and \textit{then} show that no online algorithms (including randomized algorithms) can achieve a competitive ratio smaller than $1+\ln\theta$. We formalize the two steps as follows.

\paragraph{Proof of Lemma~\ref{lem:osp-lb}.}
Let $\cala(C,v)$ denote a batch of $C$ identical buyers, each of which has a valuation $v$ ($v\in[L,U]$).
Divide the uncertainty range $[L,U]$ into a total of $m-1$ sub-ranges with equal length $\Delta v = (U-L)/(m-1)$. Let $\calv:=\{V_i\}_{i\in[m]}$ denote the $m$ boundary values with $V_i = L + (i-1) \Delta v$.  
Define an instance $I_{V_i} := \cala(C,V_{1}) \oplus \cala(C,V_2) \dots \oplus \cala(C,V_i)$, which consists of a sequence of buyer batches with increasing valuations up to $V_i$.
Here we use $\cala(C,V_i) \oplus \cala(C,V_j)$ to denote a batch $\cala(C,V_i)$ followed by a batch $\cala(C,V_j)$. 
We consider $\{I_{V_i}\}_{i\in[m]}$ as the set of hard instances for \osp.

Let $g(V_i): \calv \to [0, C]$ denote a deterministic representative function, where $g(V_i)$ is the total number of sold items under an instance $I_{V_i}$. 
Because $I_{V_{i+1}} = I_{V_i} \oplus \cala(C,V_{i+1})$ and the online decision is irrevocable, the representative function is non-decreasing, i.e., $g(V_{i+1}) \ge g(V_i), \forall i\in[m-1]$. In addition, all online algorithms must respect the capacity constraint and thus $g(V_{m}) = g(U) \le C$.
Note that each deterministic online algorithm for \osp corresponds to a unique representative function $g$, and without loss of generality, each randomized online algorithm is to randomly select a deterministic online algorithm. Let $\Tilde{g}(V_i)$ denote the total number of sold items under an instance $I_{V_i}$ by a randomized algorithm and $\bar{g}(V_i) := \ex[\Tilde{g}(V_i)]$ denote the expected representative function with respect to the randomness of the algorithm. Note that $\bar{g}$ is also non-decreasing and $\bar{g}(U) \le C$. 

Then the expected social welfare of any randomized algorithm under instance $I_{V_i}$ is 
\begin{align*}
   \ex[\alg(\Tilde{g}, I_{V_i})] = \ex\left[V_1 \Tilde{g}(V_1) + \sum\nolimits_{j=2}^i v_j [\Tilde{g}(V_j) - \Tilde{g}(V_{j-1})] \right] = L \bar{g}(L) + \sum\nolimits_{j=2}^i V_j [\bar{g}(V_j) - \bar{g}(V_{j-1})],
\end{align*} 
which can be characterized by the average representative function.
As $\Delta v \to 0$, each hard instance can be continuously indexed by $v\in[L,U]$ and the average representative function can be changed to a function of $v$. Thus, we have $\ex[\alg(\Tilde{g}, I_v)] = L \bar{g}(L) + \int_{L}^{v} u d \bar{g}(u) = v \bar{g}(v) - \int_{L}^{v} \bar{g}(u) d u$. 
In addition, the offline algorithm sells all items to the buyers in the last batch of $I_v$ and thus $\opt(I_v) = C v$.

Since any $\alpha$-competitive algorithm must satisfy $\ex[\alg(\Tilde{g}, I_v)] \ge \opt(I_v)/\alpha, \forall v\in[L,U]$, 
\begin{align}
    v \bar{g}(v) - \int_{L}^{v} \bar{g}(u) d u \ge \frac{Cv}{\alpha}, \forall v\in[L,U],
\end{align}
which is equivalent to $\bar{g}(v) \ge \frac{C}{\alpha} + \frac{1}{v} \int_{L}^{v} \bar{g}(u) d u$. Based on Gronwall's inequality (see Theorem~$1$ on Page~$356$ ~\cite{mitrinovic1991inequalities}), we have 
\begin{align}
    \label{eq:gron-ineq}
    \bar{g}(v) \ge \frac{C}{\alpha} + \frac{1}{\alpha v} \int_{L}^{v} \exp(\int_{u}^v \frac{1}{s}ds)du = \frac{C}{\alpha} + \frac{C}{\alpha} \ln\frac{v}{L}, \forall v\in[L,U].
\end{align}
Since $\bar{g}(U) \le C$, the competitive ratio is at least $\alpha \ge 1 + \ln\theta$, which gives the lower bound.  \hfill{$\square$}\smallskip

To establish the connection between the deterministic dynamic pricing algorithm (\dpa) with $C\to\infty$ and the randomized static pricing algorithm (\spa), we demonstrate that the lower bounds of the class of \dpa or the class of \spa are both $1+\ln\theta$, based on the hard instances constructed in the preceding proof. Importantly, both lower bound proofs ultimately converge to the same inequality~\eqref{eq:gron-ineq}. Here, the average function $\bar{g}$ is substituted with the deterministic function $g$ in the \dpa as $C\to\infty$, while in \spa, it is replaced with the cumulative density function of the random static threshold. Thus, inequality~\eqref{eq:gron-ineq} not only encapsulates the difficulty of the online selection problem but also characterizes the sub-classes of dynamic pricing and static pricing algorithms. Additionally, based on Gronwall's inequality, the lower bound is attained when inequality~\eqref{eq:gron-ineq} is binding for all $v\in[L,U]$, leading to the derivation of the representative function $\bar{g}^* = \frac{C}{\alpha} + \frac{C}{\alpha} \ln\frac{v}{L}, \forall v\in[L,U]$ as a byproduct. This function $\bar{g}^*$ serves as the inverse function of the pricing function $\phi$ in the \dpa and the inverse function of $\psi$ in the \spa. Consequently, the representative function that achieves the lower bound essentially guides the design of $\phi$ and $\psi$ in the posted pricing algorithms. 

{
\subsection{Static Pricing for Revenue Maximization}

Compared to \dpa, \spa achieves the optimal competitive ratio not only for social welfare maximization but also for revenue maximization. In \osp, if we use posted pricing mechanisms to maximize the revenue of the seller, i.e., $r = \sum_{n\in[N]} p_n x_n$, the offline optimal revenue is still the sum of the top $N^*$ valuations, as the seller can post prices exactly equal to the buyers' valuations. However, in the online setting, \dpa cannot attain good competitive performance since a high-valuation buyer may pay for an item at a price much lower than its valuation. For example, \dpa posts the price $L$ for the first buyer while the instance may only have one buyer with valuation $U$. In this case, the competitive ratio of \dpa for revenue maximization is $U/L = \theta$. In contrast, \spa is still $\alpha$-competitive since the expected revenue is $\int_{0}^1 \psi(x)dx = U/\alpha$. Formally, we have the following lemma.
\begin{lem}\label{lem:rev-max}
A randomized static pricing $\spa(\psi)$ with inverse CDF in equation~\eqref{eq:threshold} achieves the optimal competitive ratio $\alpha = 1+\ln\theta$ for online selection problem to maximize the revenue.     
\end{lem}
The proof of Lemma~\ref{lem:rev-max} follows directly from the proof of Theorem~\ref{thm:osp-ub}.
From the inequality~\eqref{eq:osp-ub-eq1}, we note that $\spa(\psi)$ also achieves at least ${1}/{\alpha}$ of the offline revenue, which is the same as the offline social welfare, and thus is an $\alpha$-competitive algorithm. In addition, the lower bound in Lemma~\ref{lem:osp-lb} also holds for revenue-maximization \osp.  


\smallskip 
\noindent\textbf{An application in single-leg revenue management.}
The revenue-maximization property of \spa is particularly useful in practical applications such as the single-leg revenue management problem~\cite{ball2009toward}. 
In the posted pricing setting of this problem, an airline company aims to sell $C$ seats by posting prices from a pre-determined set $\calv := \{V_1,\dots,V_m\}$, where $0< L =V_1< \dots<V_m = U$. Buyers arrive sequentially and each buyer $n$ has a valuation $v_n \in \calv$ that represents the maximum price buyer $n$ is willing to pay. The objective is to determine the posted price $p_n$ for each buyer $n$ such that the total revenue is maximized. As pointed out by prior work~\cite{ma2021policies}, deterministic dynamic pricing suffers from the risk that posting a low price might lead high-valuation buyers to substitute down. To address this, \cite{ma2021policies} proposed a randomized dynamic pricing algorithm that randomly chooses a price for each buyer to maximize revenue. However, our randomized static pricing can achieve the same goal using just one random price drawn at the beginning.

\begin{lem}\label{lem:single-leg}
A randomized static pricing algorithm $\spa(\psi)$  achieves a competitive ratio $q$ for the single-leg revenue management problem if the inverse CDF of the static price is given by
\begin{align}
\label{eq:threshold3}
\psi(x) = 
\begin{cases}
V_1 & x\in[0,Q_1]\\
V_i, & x\in (Q_{i-1}, Q_i], i=2,\dots, m,
\end{cases},
\end{align}
where $q_i = 1 - V_{i-1}/V_i, \forall i\in[m]$ with $V_0:= 0$, $q = \sum_{i=1}^m q_i$,, and $Q_i = \sum_{j=1}^i q_j/q$. $\spa(\psi)$ is an optimal online algorithm among all online algorithms.
\end{lem}
Note that the single-leg revenue management problem can be considered as a generalization of \osp because the possible posted prices are restricted to a finite set $\calv$, and the problem reduces to \osp when possible prices increase continuously from $L$ to $U$. 
However, from the technical aspect, we largely follow the proof steps of Theorem~\ref{thm:osp-ub} and Lemma~\ref{lem:osp-lb}, invoking the discrete version of Gronwall's inequality in the lower bound proof. The details are deferred to Appendix~\ref{app:proof-single-leg}.
}

\section{Online Assignment Problem}
\label{sec:omsp}

We continue studying the online assignment problem (\ogap), which is an extension of the \osp. In this problem, a seller maintains $K$ items, with each item $k$ having $C_k$ copies for sale. $N$ buyers arrive sequentially. Upon the arrival of each buyer $n\in[N]$, the seller posts a price $p_{n,k}$ for each item $k\in[K]$.
Buyer $n$ has private valuations over the items $\{v_{n,k}\}_{k\in[K]}$, where $v_{n,k}$ represents her valuation for one unit of item $k$. Without loss of generality, $v_{n,k} = 0$ if buyer $n$ is not interested in item $k$. Based on the posted prices, buyer $n$ can obtain a utility of $v_{n,k} - p_{n,k}$ from purchasing item $k$. Then buyer $n$ decides to purchase an item if the item achieves the maximum utility and the maximum utility is non-negative.
Let $x_{n,k} \in \{0,1\}$ denote whether buyer $n$ purchases item $k$. Buyer $n$ obtains a utility of $u_n = \sum_{k\in[K]} x_{n,k}(v_{n,k} - p_{n,k})$, and the seller collects a total revenue of $r_k = \sum_{n\in[N]} x_{n,k} p_{n,k}$ from selling item $k$.
The goal of the problem is to maximize the social welfare of all buyers and the seller, i.e., $\sum_{n\in[N]} u_n + \sum_{k\in[K]} r_k = \sum_{n\in[N]}\sum_{k\in[K]} x_{n,k} v_{n,k}$.

We still use $I :=\{v_{n,k}\}_{n\in[N],k\in[K]}$ to denote an instance of \ogap. Given the instance, the optimal social welfare $\opt(I)$ can be obtained by solving the following offline problem: 
\begin{subequations}
\label{p:gap}
\begin{align}
    \max_{x_{n,k}} \quad& \sum\nolimits_{n\in[N]}\sum\nolimits_{k\in[K]} v_{n,k} x_{n,k}\\
    \label{eq:oap-eq1}
    {\rm s.t.}\quad& \sum\nolimits_{n\in[N]}  x_{n,k} \le C_k, \forall k\in[K],\\
    \label{eq:oap-eq2}
    &\sum\nolimits_{k\in[K]} x_{n,k} \le 1, \forall n\in[N],\\
    & x_{n,k}\in\{0,1\}, \forall n\in[N],k\in[K],
\end{align}    
\end{subequations}
where constraint~\eqref{eq:oap-eq1} guarantees that at most $C_k$ copies of item $k$ are sold and constraint~\eqref{eq:oap-eq2} ensures that each buyer purchases at most one item.
In the online setting, a posted price algorithm must determine the price vector $\{p_{n,k}\}_{k\in[K]}$ for each buyer $n$ just based on past purchase decisions $\{x_{i,k}\}_{k\in[K], i< n}$ without the information of future buyers. We aim to find the online pricing algorithm that can achieve the minimum competitive ratio. 

The \ogap problem is closely related to many online optimization/decision problems. Specifically, it can be viewed as an online edge-weighted matching problem, where offline vertices correspond to items and online vertices correspond to buyers in \ogap, and each matching between an offline node $k$ and an online node $n$ generates a reward $v_{n,k}$. It is well-known that no online algorithm can achieve a bounded competitive ratio in the general online edge-weighted matching problem~\cite{mehta2013online}. Therefore, prior work relies on various additional assumptions based on specific applications. For example, this problem has been studied under the free disposal assumption in the Ad assignment problem~\cite{feldman2009online}. In this paper, we extend the bounded valuation assumption of \osp to \ogap.

\begin{ass}\label{ass:bounded-density3}
For each item $k\in[K]$, buyers' valuations are bounded if they are interested in item $k$, i.e., $v_{n,k}\in[L_k,U_k], \forall n\in[N]$.
\end{ass}
Let $\theta_k = U_k/L_k$ denote the fluctuation ratio of item $k$, and let $\theta = \max_{k\in[K]} \theta_k$. Under Assumption~\ref{ass:bounded-density3}, prior work~\cite{ma2020algorithms} has proposed a dynamic pricing algorithm for the \ogap. For each item $k$, the posted price is determined by a pricing function $\phi_{k}(y): [C_k] \to [L_k, U_k]$, where $\phi_{k}(y)$ is the posted price for the $y$-th unit of item $k$. In the large supply regime, where $\min_{k\in[K]} C_k \to \infty$, \cite{ma2020algorithms} designs an optimal pricing function for dynamic pricing algorithms. By combining the multi-price balance algorithm, Theorem 1, and Appendix E.1 in~\cite{ma2020algorithms}, we can have the following lemma.
\begin{lem}[\cite{ma2020algorithms}]\label{lem:oap-up}
There exists a dynamic pricing algorithm that can achieve a competitive ratio of $\alpha_{\ogap}^{\infty}$ when the posted prices are determined by pricing functions, $\forall k\in[K]$,
\begin{align}\label{eq:thred-ogap}
\phi_k(y) = 
\begin{cases}
    L_k \cdot \frac{e^{y/C_k} - 1}{e^{\omega_k} - 1} & y \in [0,\omega_k\cdot C_k)\\
    \frac{U_k - L_k}{e^{\alpha_k} - e^{\omega_k \cdot \alpha_k}} e^{\alpha_k y/C_k} &y\in [\omega_k \cdot C_k, C_k]
\end{cases},
\end{align}
where $\omega_k$ is the solution of $\frac{e^\omega}{e^\omega - 1} = \frac{\ln\theta_k}{1 -\omega}$, $\alpha_k = \frac{e^{\omega_k}}{e^{\omega_k} - 1}$ and $\alpha_{\ogap}^{\infty} = \max_{k\in[K]} \alpha_k$.
\end{lem}
Since the original algorithm and results are not presented under the posted pricing mechanism, for the sake of completeness, we provide the detailed dynamic pricing algorithm and proof of Lemma~\ref{lem:oap-up} in Appendix~\ref{app:lem-oap-up}. In the following section, we focus on the design and analysis of a randomized static pricing algorithm that can achieve competitive results comparable to those of dynamic pricing.

\subsection{Static Pricing for Online Assignment Problem}

We propose a randomized static pricing algorithm for \ogap. This algorithm determines a static price $p_k$ for each item $k$ by sampling based on an inverse CDF distribution $\psi_k$, and then keeps using the prices until the end of the algorithm. Upon the arrival of each buyer $n$, the seller posts the fixed price vector $\bp:=\{p_k\}_{k\in[K]}$. The buyer then determines the item $k^*$ that maximizes her utility $\max_{k\in[K]} v_{n,k} - p_k$. Then the buyer purchases $k^*$ if the resulting utility $v_{n,k^*} - p_{k^*}$ is non-negative and rejects all prices otherwise. 
See Algorithm~\ref{alg:spa-oap} in Appendix~\ref{app:oap-ub} for more detail.
The static pricing algorithm is fully parameterized by the inverse CDF function $\psi:=\{\psi_k\}_{k\in[K]}$. 
We still use $\spa(\psi)$ to denote the static pricing algorithm.
By carefully designing $\psi$, $\spa(\psi)$ can attain the same competitive ratio as the dynamic pricing in the large supply regime.

\begin{thm}\label{thm:oap-ub}
A randomized static pricing algorithm $\spa(\psi)$ is $\alpha_{\ogap}^{\infty}$-competitive for the online assignment problem when the inverse CDF function $\psi:=\{\psi_k\}_{k\in[K]}$ is given by $\forall k\in[K]$,
    \begin{align}\label{eq:thred-oap}
        \psi_k(x) = 
        \begin{cases}
            \frac{(\alpha_k - 1)L_k}{\alpha_k} \cdot e^{x} & x \in [0,\omega_k)\\
            U_k e^{\alpha_k (x - 1)} &x\in [\omega_k, 1]
        \end{cases},
    \end{align}
where $\omega_k$ is the solution of $\frac{e^\omega}{e^\omega - 1} = \frac{\ln\theta_k}{1 -\omega}$, $\alpha_k = \frac{e^{\omega_k}}{e^{\omega_k} - 1}$ and $\alpha_{\ogap}^{\infty} = \max_{k\in[K]} \alpha_k$.
\end{thm}

We extend the economics-based approach of \osp to analyze $\spa(\psi)$.  
Different from the analysis in \osp, a buyer $n$ in \ogap may choose not to purchase item $k$ even if her valuation $v_{n,k}$ is higher than the posted price $p_k$ of item $k$. This is because \spa may instead select another item that can provide greater utility. Such competition among items leads to the coupling of the online decision for one item and the posted prices of the other items.
To capture the influence of the posted prices from other items, we define an \textit{effective valuation} $\hat{v}_{n,k}$ for each buyer $n$ and item $k$ and use $\hat{v}_{n,k}$ to quantify the impact of other posted prices on the purchase decision of item $k$.

In particular, for each instance $I$, let $I^{(n,k)}$ denote a modified instance, in which the valuation $v_{n,k}$ of buyer $n$ for item $k$ is set to $0$, forcing buyer $n$ not to purchase item $k$.
For a given posted price $\bp$, let $u^{(n,k)}_n$ denote the utility of buyer $n$ from \spa under the modified instance $I^{(n,k)}$. We call $u^{(n,k)}_n$ a shadow utility, which captures the competition from other items when buyer $n$ considers purchasing item $k$. The larger $u^{(n,k)}_n$ is, the more likely buyer $n$ is to purchase items other than $k$. Then we define the effective valuation of buyer $n$ on item $k$ as
\begin{align*}
 \hat{v}_{n,k} := v_{n,k} - \min\{u^{(n,k)}_n, v_{n,k}\}.   
\end{align*}
The shadow utility $u^{(n,k)}_n$ and effective valuation $\hat{v}_{n,k}$ satisfy the following properties: (i) The utility of buyer $n$ is not less than its shadow utility, i.e., $u_n \ge u^{(n,k)}_n$; and (ii) Buyer $n$ purchases item $k$ if her effective valuation exceeds the posted price $p_k$, i.e., $\hat{v}_{n,k} > p_k$, and the item $k$ has not been sold out.
With these two properties, we can analyze the expected revenue from item $k$ using similar arguments in the proof of \spa for \osp, by replacing the valuations of buyers with the effective valuations.
The proof details are given in Appendix~\ref{app:oap-ub}.

\subsection{Lower Bound for Online Assignment Problem}

This section provides a lower bound for \ogap, which matches the competitive ratio $\alpha_{\ogap}^{\infty}$ attained by the dynamic pricing in the large supply regime and the proposed static pricing in a general supply regime. Thus, our proposed static pricing algorithm is optimal among all online algorithms. 

\begin{thm}
\label{thm:lb-oap}
No online algorithm, including deterministic and randomized algorithms, for the online assignment selection problem can achieve a competitive ratio smaller than $\frac{e^\omega}{e^\omega - 1}$, where $\omega$ is the solution of $\frac{e^\omega}{e^\omega - 1} = \frac{\ln\theta}{1 -\omega}$.
\end{thm}

We still rely on a representative function approach to demonstrate the lower bound of $\ogap$. We first construct a family of hard instances, where each instance can be divided into two stages. The instance in Stage I is the classic upper-triangle instance from the online matching literature~\cite{devanur2012online}, which requires the online algorithm to balance the numbers of sold units from different items. The Stage II of the instance follows the design of the  worst-case instance for the online selection problem constructed in the proof of Lemma~\ref{lem:osp-lb}, which requires the online algorithm to reserve some units for high-valuation buyers. Because of the possible occurrence of the Stage II, algorithm cannot fully utilize its inventory for each of the items in the first stage. 
Let $\omega$ denote the maximum utilization reserved for items in Stage I, the competitive ratio of any algorithms is lower bounded by $\alpha \ge \frac{e^\omega}{e^\omega - 1}$, following similar arguments in the lower bound proof of online matching. 
At the end of Stage I, the arrivals of Stage II are constructed similarly to the worst-case instance of the online selection problem in the proof of Lemma~\ref{lem:osp-lb}, and only interested in one of the items that have been sold up to $\omega$. We modify the lower bound proof of \osp by additionally taking into account the selection decisions in Stage I, and lower bound the competitive ratio by $\alpha \ge \frac{\ln\theta}{1-\omega}$.  
The lower bound of \ogap is obtained by optimizing the representative function to balance the difficulties from the instances in the two stages. The proof details are presented in Appendix~\ref{app:oap-lb}.

\section{Online Selection with Convex Costs}
\label{sec:oscc}

In this section, we consider another variant of the online selection problem that considers the cost of producing the items.
Specifically, a seller aims to sell $C$ units of an item to $N$ buyers that arrive sequentially. 
The $C$ units of items are produced at a non-decreasing marginal cost. Let $f(i)$ denote the cost of producing the first $i$ units, which is a convex function.
Upon the arrival of buyer $n$, the seller posts a price $p_n$, and the buyer $n$ decides to purchase one item if her valuation is greater than the price. Let $x_n \in\{0,1\}$ denote the decision of buyer $n$ that determines whether to purchase the item.
Buyer $n$ can obtain a utility $u_n = x_n (v_n - p_n)$ and the seller can collect a total profit of $r = \sum_{n\in[N]} x_n p_n - f(\sum_{n\in[N]}x_n)$.
The goal of the online selection with convex cost (\oscc) is to maximize the social welfare of all buyers and the seller, i.e., $\sum_{n\in[N]}u_n + r = \sum_{n\in[N]} v_n x_n - f(\sum_{n\in[N]}x_n)$.
Given an arrival instance $I = \{v_n\}_{n\in[N]}$ of \oscc, the offline problem can be formulated as
\begin{align}
\label{p:okpc-primal}
\max_{x_n} \quad \sum\nolimits_{n\in[N]} v_n x_n - f\left(\sum\nolimits_{n\in[N]} x_n \right), \quad {\rm s.t.} \sum\nolimits_{n\in[N]} x_n \le C, x_n \in \{0,1\}, \forall n\in[N].
\end{align}
In \oscc, we still assume that the buyers' valuations are bounded within $[L,U]$ as in Assumption~\ref{ass:bounded-value}.
In addition, the production cost function $f(i): \calc\to\mathbb{R}^+$ is a non-decreasing convex function, where $\calc:=\{0,1,\dots,C\}$. Then the marginal production cost $c_i = f(i) - f(i-1)$ is non-negative and non-decreasing in $i$. 
We consider the supply cost function with zero setup cost, i.e., $f(0) = 0$.

Before proceeding to the detailed algorithm, we introduce some definitions and notations.
For a given valuation $v\in[L, U]$, define the conjugate function of the production cost function $f$ as
\begin{align}
    h(v) = \max_{y\in\calc}\quad v y - f(v), 
\end{align}
which can be interpreted as the maximum profit a seller can obtain when the instance only contains $C$ buyers with identical valuation $v$. 
Let $y^{\texttt{inv}}(v) = \arg\max_{y\in\calc} v y - f(v)$ denote the optimal solution of the above optimization problem for a given $v$.
Then $\bar{C} := y^{\texttt{inv}}(U)$ can denote the \textit{effective capacity} of \oscc since the marginal cost of the $(\bar{C} + 1)$-th item is larger than the maximum possible valuation $U$ and thus there is no incentive to produce more than $\bar{C}$ units of the item.

Prior work has designed dynamic pricing algorithms for \oscc in both the large supply regime~\cite{tan2020mechanism} and the finite supply regime~\cite{tan2023threshold}. The pricing functions and the corresponding competitive ratios are characterized as solutions of differential equations and exhibit no analytical forms in general. We use $\alpha_{\oscc}^{\infty}$ and $\alpha_{\oscc}^C$ to denote the optimal competitive ratios in the large supply and finite supply regimes, respectively.

\subsection{Static Pricing for Online Selection with Convex Cost}

We can use the static pricing algorithm (\spa) described in Algorithm~\ref{alg:osp-spa} to solve \oscc by replacing the capacity $C$ in line~\ref{alg:line-capacity} with the effective capacity $\bar{C}$.
We first show a lower bound for \spa when the static price is set deterministically. 

\begin{lem}
No deterministic static pricing algorithm can achieve a competitive ratio smaller than $\frac{h(U)}{h(L)}$ for the online selection problem with convex cost.
\end{lem} 
To show this, we can first claim that any static price above $L$ results in unbounded competitive ratios. Thus, the static price must be set as $L$. Then we can construct an instance, in which the buyers' valuations increase from $L$ to $U$ and the total number of buyers with the same valuation is $C$. The worst-case ratio of the static pricing algorithm can be shown to be $\frac{h(U)}{h(L)}$ under this instance.  

Thus, we focus on the randomized static pricing algorithm $\spa(\psi)$. We next show that randomization can improve the competitive ratio to $O(\ln\frac{h(U)}{h(L)})$. 

\begin{thm}
\label{thm:oscc-ub}
For the online selection with convex cost, the static pricing algorithm $\spa(\psi)$ can achieve a competitive ratio $\alpha_{\oscc}^{\texttt{sta}}$ if $\psi = G^{-1}$ with the CDF of the static price 
\begin{align}\label{eq:oscc-dist}
  G(v) = \frac{1}{\alpha} + \frac{1}{\alpha} \cdot \ln\frac{h(v)}{h(L)}, \quad v\in [L,U],
\end{align}
where $\alpha = \alpha_{\oscc}^{\texttt{sta}} = 1 + \ln\frac{h(U)}{h(L)}$ and $h(\cdot)$ is the conjugate function of the production cost $f(\cdot)$.
\end{thm}
{
In the following we make two remarks on Theorem \ref{thm:oscc-ub} and defer the formal proof to Appendix~\ref{app:oscc-ub}. First, in the next section, it can be shown that $\alpha_{\oscc}^{\texttt{sta}}$ is, in fact, the best possible competitive ratio among all randomized static pricing algorithms. Second, \cite{tan2020mechanism} has shown that $\alpha_{\oscc}^{\infty}$ from dynamic pricing cannot be characterized in closed form, and thus it is difficult to compare $\alpha_{\oscc}^{\texttt{sta}}$ and $\alpha_{\oscc}^{\infty}$ in general. Nevertheless, we can make some interesting observations in certain special cases. Particularly, in a high-valuation case, where the maximum marginal production cost $f(C) - f(C-1)$ is smaller than the buyers' valuation lower bound $L$, the competitive ratios of static pricing and dynamic pricing can be compared as follows
\begin{align}
   \alpha_{\oscc}^{\texttt{sta}} = 1 + \ln\frac{U - f(C)/C}{L - f(C)/C} \ge 1 + \ln\theta \ge \alpha_{\oscc}^{\infty},
\end{align}
where $\alpha_{\oscc}^{\texttt{sta}}$ is obtained by observing $h(v) = v C - f(C), \forall v\in[L,U]$ in this case, and $\alpha_{\oscc}^{\infty} \le 1 + \ln\theta$ is the result from~\cite{tan2023threshold}. 
We can observe that $\alpha_{\oscc}^{\texttt{sta}}$ is worse than the optimal competitive ratio for online selection without production cost, which is further worse than $\alpha_{\oscc}^{\infty}$---the optimal competitive ratio of dynamic pricing in the large supply setting. Moreover, the competitive ratio $\alpha_{\oscc}^{\texttt{sta}}$ becomes larger as the production cost increases faster (i.e., $f(C)/C$ gets larger). This is in contrast to $\alpha_{\oscc}^{\infty}$, which is shown to decrease as the production cost increases faster~\cite{tan2023threshold}. Thus, the convexity of the production cost intricately limits the performance of static pricing (in terms of competitive ratios) and, conversely, enhances the performance of dynamic pricing. This can be interpreted as the inherent price a seller must pay to avoid price discrimination. 
}
 

We can further show that the proposed algorithm in Theorem~\ref{thm:oscc-ub} attains the best possible competitive ratio among all static pricing algorithms. 

\begin{thm}
\label{thm:oscc-lb}
For the online selection with convex cost, no randomized static pricing algorithm can achieve a competitive ratio smaller than $1+\ln\frac{h(U)}{h(L)}$.
\end{thm}
Theorem~\ref{thm:oscc-lb} is proved based on the same hard instances constructed in Section~\ref{sec:osp-lb} in the proof of \osp while this proof focuses on the randomized static pricing algorithm using the CDF of the static price instead of the representative function. The full proof is given in Appendix~\ref{app:oscc-lb}.

\section{Conclusions and Future Directions}
In this paper, we have designed and analyzed static posted pricing algorithms for the adversarial online selection problem and its two important variants: the online assignment problem and the online selection with convex cost. Compared to dynamic pricing, static pricing algorithms are simple to implement and have the merit of avoiding price discrimination. Previous studies in the context of stochastic online selection (e.g., the prophet inequality problem) have shown that, in general, static pricing is inferior to dynamic pricing  in social welfare or revenue maximization. Our results show that simple static pricing algorithms can achieve surprisingly strong guarantees comparable to the best possible dynamic pricing algorithms for the adversarial online selection problem. To achieve this result, we adopt an economics-based approach in the competitive analysis of static pricing algorithms and propose a novel representative function-based proof to establish the lower bound of the adversarial online selection problem and its two variants. We expect that our proof techniques will also be useful in related online problems such as online matching.

Our work motivates several interesting new problems, including: (i) the design and analysis of static pricing algorithms for online combinatorial auctions and their variants; (ii) extension to a reusable resource setting, where each item can be rented for a duration instead of being sold; (iii) studying the risk sensitivity of the randomized static pricing algorithms to go beyond the current risk-neutral analysis based on the expected performance of the randomized algorithm.

\newpage
\appendix
\clearpage
\section*{Acknowledgements}
Bo Sun and Raouf Boutaba acknowledge the support from the NSERC Discovery Grant RGPIN-2019-06587. Xiaoqi Tan acknowledges support from Alberta Machine Intelligence Institute (Amii), Alberta Major Innovation Fund, and NSERC Discovery Grant RGPIN-2022-03646.
\bibliography{reference}
\bibliographystyle{plain}

\clearpage
\appendix
\section*{\centering\LARGE{Appendix}}

\section{Proofs for Online Selection Problem (\osp)}

\subsection{Proof of Lemma~\ref{lem:ub-osp-dp}}
\label{app:lem-osp-dp}
We use a constructive approach to demonstrate how to design dynamic posted prices that can minimize the competitive ratio. 
Let $\Phi := \{\Phi_i\}_{i\in[C]}$ denote the $C$ posted prices in the pricing function $\phi_{\texttt{OSP}}^C$, where $\Phi_i = \phi_{\texttt{OSP}}^C(i)$ represents the price for the $i$-th unit of the item.

To maximize social welfare, the seller needs to aggressively sell items at the beginning to hedge the risk that no buyers may come in the future. As more items are sold, the algorithm can gradually become more selective by posting higher prices, taking the opportunity to sell at potentially high prices. Thus, we can focus on the posted prices that are monotonically non-decreasing, and the first $\gamma \in [C]$ prices are set to the lowest price $L$, i.e., $\Phi_i = L$ for all $i=1, \dots, \gamma$. 

Let $I = \{v_1,\dots,v_N\}$ denote an instance of the online selection problem, and let $\opt(I)$ and $\alg(\Phi,I)$ denote the social welfare of the offline algorithm and the dynamic pricing algorithm, respectively. Let $I'$ denote an instance that contains the same set of buyers as $I$ but arranges the buyers' valuations in a non-decreasing order. It is easy to verify that $\opt(I) = \opt(I')$ and $\alg(\Phi,I) \ge \alg(\Phi,I')$. Consequently, without loss of generality, we can focus on the instances in which the buyers' valuations are non-decreasing over time.

Given an instance $I$, let $z$ denote the number of sold units under the dynamic pricing algorithm with posted prices $\Phi$. If $z < \gamma$, it indicates that the total number of buyers $N = z$, since the algorithm keeps posting the lowest possible price $L$ before selling $\gamma$ units.

If $z \ge \gamma$, the social welfare of the offline algorithm is upper-bounded by $\opt(I) \le \Phi_{z+1} C$, and the social welfare of the dynamic pricing algorithm is at least $\alg(\Phi,I) \ge \sum_{i\in[z]}\Phi_i = \gamma L + \sum_{i=\gamma+1}^z \Phi_i$. Thus, to ensure $\alpha$-competitiveness, the posted prices must be designed such that:
\begin{align*}
    \frac{C L}{\gamma L} &\le \alpha,\\
    \frac{C \Phi_{z+1}}{\gamma L + \sum_{i=\gamma +1 }^z \Phi_i } &\le \alpha, z = \gamma, \gamma+1,\dots, C.
\end{align*}
The dynamic prices $\Phi$ (given in equation~\eqref{eq:threshold-osp}) are designed to satisfy the above inequalities and minimize the corresponding competitive ratio.

From the construction of the dynamic posted prices, we see that in fact no deterministic online algorithm can attain a better competitive ratio in the worst case.

\subsection{Proof of Lemma~\ref{lem:single-leg}}
\label{app:proof-single-leg}
The upper bound proof follows exactly the same steps as that of Theorem~\ref{thm:osp-ub} until equation~\eqref{eq:ineq2}. Then we can continue to lower bound the expected revenue of $\spa(\psi)$ by
\begin{subequations}
\begin{align}
\mathbb{E}[\alg(p,I)] &\ge  N^*\int_{0}^{x_1}  \psi(x) dx + \sum_{i=2}^{N^*} \int_{x_{1}}^{x_{i}} \psi(x)dx \\
\label{eq:ineq4}
& \ge  N^*\frac{v_1^*}{q} + \sum_{i=2}^{N^*} \frac{v_i^* - v_1^*}{q}\\
&= \frac{1}{q} \sum_{i\in[N^*]} v_i^* = \frac{1}{q} \opt(I).
\end{align}    
\end{subequations}
Let $\kappa(i)$ denote the index of the price such that $V_{\kappa(i)} = v_i^* = \psi(x_i)$.
The inequality~\eqref{eq:ineq4} holds since the proposed inverse CDF $\psi$ in equation~\eqref{eq:threshold1} gives
\begin{subequations}
\begin{align}
    \int_{0}^{x_1} \psi(x)dx &= \sum_{j=1}^{\kappa(1)} [Q_j- Q_{j-1}]V_j = \frac{1}{q} \sum_{j=1}^{\kappa(1)} [V_j - V_{j-1}] = \frac{V_{\kappa(1)}}{q} = \frac{v_1^*}{q}, \\
    \int_{x_1}^{x_i} \psi(x) dx &= \sum_{j=\kappa(1)}^{\kappa(i)} [Q_j- Q_{j-1}]V_j = \frac{1}{q} \sum_{j=\kappa(1)}^{\kappa(i)} [V_j - V_{j-1}]  = \frac{v_i^* - v_1^*}{q}, \quad i = 2,\dots, N^*. 
\end{align}    
\end{subequations}

We use $\calv$ to denote the set of predetermined price set instead of the boundaries of the equally discretized intervals. The lower bound proof follows that of Lemma~\ref{lem:osp-lb} until we can obtain the expected revenue of any randomized algorithm under instance $I_{V_i}$ by 
\begin{align*}
   \ex[\alg(\Tilde{g}, I_{V_i})] = V_1 \bar{g}(V_1) + \sum_{j=2}^i V_j [\bar{g}(V_j) - \bar{g}(V_{j-1})] = V_i \bar{g}(V_i) - \sum_{j=2}^{i} \bar{g}(V_{j-1}) [V_{j} - V_{j-1}],
\end{align*}

Since any $q$-competitive algorithm must satisfy $\ex[\alg(\Tilde{g}, I_{V_i})] \ge \opt(I_{V_i})/q, \forall i\in[m]$, 
\begin{align*}
    V_i \bar{g}(V_i) - \sum_{j=2}^{i} \bar{g}(V_{j-1}) [V_{j} - V_{j-1}] \ge \frac{C \cdot V_i}{q},
\end{align*}
which is equivalent to, 
\begin{subequations}
\label{eq:gron-ineq-discrete}   
\begin{align}
\bar{g}(V_i) &\ge \frac{C}{q} + \frac{1}{V_i} \sum_{j=2}^{i} \bar{g}(V_{j-1}) [V_{j} - V_{j-1}]\\
&\ge \frac{C}{q} + \frac{C}{q V_i} \sum_{j=2}^{i} \left[(V_{j} - V_{j-1})\cdot \prod_{k = j}^{i} \left(1 + \frac{V_{k} - V_{k-1}}{V_{k-1}} \right) \right]\\
&= \frac{C}{q} + \frac{C}{q} \sum_{j=2}^{i} \left[1 - \frac{V_{j-1}}{V_{j}} \right],
\end{align}
\end{subequations}
where the second inequality is based on a discrete version of Gronwall's inequality. 
Since $\bar{g}(V_m) \le C$, the competitive ratio is at least
\begin{align}
    q \ge 1 + \sum_{j=2}^{m} \left[1 - \frac{V_{j-1}}{V_{j}} \right].
\end{align}
$\quad\square$

\section{Proofs for Online Assignment Problem (\ogap)}

\subsection{Proof of Lemma~\ref{lem:oap-up}}
\label{app:lem-oap-up}

We describe the dynamic pricing algorithm with pricing function $\phi$ ($\dpa(\phi)$) in Algorithm~\ref{alg:dpa-oap} for the online assignment problem. The main difference between $\spa(\psi)$ and $\dpa(\phi)$ is that $\dpa(\phi)$ determines the posted prices as a function of the number of sold units (i.e., $y_k$ in the algorithm). We next analyze the competitive ratio of the $\dpa(\phi)$ based on an online primal-dual approach.

The relaxed primal and dual of the offline problem~\eqref{p:gap} can be described as
\begin{subequations}
\begin{align}
   (\texttt{Primal})\quad \max_{x_{n,k} \ge 0} \quad& \sum_{n\in[N]}\sum_{k\in[K]} v_{n,k} x_{n,k}\\
    {\rm s.t.}\quad& \sum_{n\in[N]}  v_{n,k} \le C_k, \forall k\in[K], \quad (\lambda_k)\\
    &\sum_{k\in[K]} x_{n,k} \le 1, \forall n\in[N]. \quad (\eta_n)
\end{align}    
\end{subequations}
\begin{subequations}
\begin{align}
   (\texttt{Dual})\quad  \min_{\lambda_k \ge 0, \eta_n \ge 0} \quad& \sum_{k\in[K]} \lambda_{k} C_k + \sum_{n\in[N]} \eta_{n}\\
    {\rm s.t.}\quad& \eta_n \ge v_{n,k} - \lambda_k, \forall n\in[N], k\in[K].
\end{align}    
\end{subequations}

Let $\{\bar{x}_{n,k}\}_{n\in[N],k\in[K]}$ denote the online solution of $\dpa(\phi)$. Let $y_k^{(n)} = \sum_{m=1}^n \bar{x}_{m,k}$ denote the number of sold units of item $k$ to the first $n$ buyers. 
We can construct dual variables based on the online solutions as
\begin{subequations}
\begin{align}
    \bar{\lambda}_k &= \phi_k(y_k^{(N)}), \forall k\in[K],\\
    \bar{\eta}_n &= \sum_{k\in[K]} \bar{x}_{n,k} [v_{n,k} - \phi_k(y_k^{(n-1)})], \forall n\in[N].
\end{align}    
\end{subequations}
It is clear that the primal and dual variables are both feasible. Based on weak duality, we have
\begin{subequations}
\begin{align}
    \opt(I) &\le \sum_{k\in[K]} \bar{\lambda}_{k} C_k + \sum_{n\in[N]} \bar{\eta}_{n}\\
    &= \sum_{k\in[K]} \phi_k(y_k^{(N)}) C_k + \sum_{n\in[N]} \sum_{k\in[K]} \bar{x}_{n,k} [v_{n,k} - \phi_k(y_k^{(n-1)})]\\
    &= \sum_{n\in[N]} \sum_{k\in[K]} \bar{x}_{n,k} v_{n,k} + \sum_{k\in[K]} \left[\phi_k(y_k^{(N)}) C_k - \sum_{n\in[N]} \bar{x}_{n,k}\phi_k(y_k^{(n-1)})\right]\\
    \label{eq:approx}
    &\approx \sum_{n\in[N]} \sum_{k\in[K]} \bar{x}_{n,k} v_{n,k} + \sum_{k\in[K]} \left[\phi_k(y_k^{(N)}) C_k - \sum_{n\in[N]} \int_{y_k^{(n-1)}}^{y_k^{(n)}} \phi_k(u)du\right]\\
    &=\sum_{n\in[N]} \sum_{k\in[K]} \bar{x}_{n,k} v_{n,k} + \sum_{k\in[K]} \left[\phi_k(y_k^{(N)}) C_k - \int_{0}^{y_k^{(N)}} \phi_k(u)du\right]\\
    \label{eq:crit-ineq}
    &\le \sum_{n\in[N]} \sum_{k\in[K]} \bar{x}_{n,k} v_{n,k} + \sum_{k\in[K]}(\alpha_k - 1)\sum_{n\in[N]}  \bar{x}_{n,k} v_{n,k} \\
    &\le \alpha \alg,
\end{align}    
\end{subequations}
where $\alpha = \max_{k\in[K]} \alpha_k$.
In the large supply regime, we have $\int_{y_k^{(n-1)}}^{y_k^{(n)}} \phi_k(u)du \approx \bar{x}_{n,k}\phi_k(y_k^{(n-1)})$ and thus the approximation~\eqref{eq:approx} holds. 
To show the inequality~\eqref{eq:crit-ineq}, we first note the designed pricing function~\eqref{eq:thred-ogap} can ensure the following inequality, $\forall k\in[K]$,
\begin{subequations}
\begin{align*}
   &\phi_k(y)C_k - \int_{0}^y \phi_k(z)dz \le (\alpha_k - 1) L_k \cdot y, \quad y\in[0, \omega_k \cdot C_k),\\
   & \phi_k(y)C_k - \int_{0}^y \phi_k(z)dz \le (\alpha_k - 1) [\omega_k C_k L_k + \int_{\omega_k C_k}^y \phi_k(z) dz], \quad y\in[\omega_k \cdot C_k, C_k].
\end{align*}    
\end{subequations}
The inequality~\eqref{eq:crit-ineq} holds since 
\begin{align*}
   \phi_k(y_k^{(N)}) C_k - \int_{0}^{y_k^{(N)}} \phi_k(u)du \le  (\alpha_k - 1) \sum_{n\in[N]} \bar{x}_{n,k} v_{n,k}.
\end{align*}
To see the above equation, we make the following observations.

(i) When $y_k^{(N)} < \omega_k C_k$,  
\begin{align*}
  \phi_k(y_k^{(N)}) C_k - \int_{0}^{y_k^{(N)}} \phi_k(u)du \le  (\alpha_k - 1) y_k^{(N)} L_k\le (\alpha_k - 1)\sum_{n\in[N]} \bar{x}_{n,k} v_{n,k}.
\end{align*}
 
(ii) When $\omega_k C_k \le y_k^{(N)} \le C_k$, we have
\begin{align*}
    \phi_k(y_k^{(N)}) C_k - \int_{0}^{y_k^{(N)}} \phi_k(u)du &\le (\alpha_k - 1) [\omega_k C_k L_k + \int_{\omega_k C_k}^y \phi_k(z) dz]\\
    &\le(\alpha_k - 1) \sum_{n\in[N]} \bar{x}_{n,k} v_{n,k}.
\end{align*}

Thus, the competitive ratio of $\dpa(\phi)$ is $\alpha = \max_{k\in[K]} \alpha_k$. This completes the proof.

\begin{algorithm}[t]
\caption{Dynamic Price Algorithm for Online Assignment Problem ($\dpa(\phi)$)}
\label{alg:dpa-oap}
\begin{algorithmic}[1]
\State \textbf{input:} pricing function $\phi:=\{\phi_k(\cdot)\}$; 
\State \textbf{initiate:} $y_k = 1, \forall k\in[K]$; 
\While{$n = 1,\dots,N$} \Comment{buyer $n$ arrives with valuation $\{v_{n,k}\}_{k\in[K]}$}
\State post price $p_{n,k} = \phi_k(y_k)$ for item $k$, $\forall k\in[K]$; 
\State determine $k^* = \max_{k\in[K]} v_{n,k} - p_{n,k}$; \Comment{determine utility-maximizing item}
\If{$v_{n,k^*} - p_{n,k^*} \ge 0$ and $y_{k^*} \le C_k$}
\State set $x_{n,k^*} = 1$ and  $x_{n,k} = 0, \forall k\not= k^*$; \Comment{buyer $n$ accepts the price of item $k$}
\State update $y_{k^*} = y_{k^*} + 1$;
\Else
\State set $x_{n,k} = 0, \forall k \in [K]$; \Comment{buyer $n$ rejects all prices}
\EndIf
\EndWhile
\end{algorithmic}
\end{algorithm}

\subsection{Proof of Theorem~\ref{thm:oap-ub}}
\label{app:oap-ub}

\begin{algorithm}[h]
\caption{Static Pricing Algorithm for Online Assignment Problem}
\label{alg:spa-oap}
\begin{algorithmic}[1]
\State \textbf{input:} inverse cumulative density function $\psi:=\{\psi_k(\cdot)\}$; 
\State sample a uniform random variable $x_k$ within $[0,1]$ {independently} for each $k\in[K]$;
\State \textbf{initiate:} $z_k = 1, \forall k\in[K]$; 
\While{$n = 1,\dots,N$} \Comment{buyer $n$ arrives with valuation $\{v_{n,k}\}_{k\in[K]}$}
\State post price $p_k = \psi_k(x_k)$, $\forall k\in[K]$; 
\State determine $k^* = \max_{k\in[K]} v_{n,k} - p_k$; \Comment{determine utility-maximizing item}
\If{$v_{n,k^*} - p_{k^*} \ge 0$ and $z_{k^*} \le C_k$}
\State set $x_{n,k^*} = 1$ and  $x_{n,k} = 0, \forall k\not= k^*$; \Comment{buyer $n$ accepts the price of item $k$}
\State update $z_{k^*} = z_{k^*} + 1$;
\Else
\State set $x_{n,k} = 0, \forall k \in [K]$; \Comment{buyer $n$ rejects all prices}
\EndIf
\EndWhile
\end{algorithmic}
\end{algorithm}

Given an instance $I:=\{v_{n,k}\}_{n\in[N],k\in[K]}$ of the \ogap problem and a realization of $K$ posted prices $\bp = \{p_k\}_{k\in[K]}$, let $\{X_{n,k}\}_{n\in[N],k\in[K]}$ denote the online solution of $\spa$. Then the utility of buyer $n$ and the total revenue earned by selling item $k$ can be determined by
\begin{align*}
    u_{n} &:= \sum\nolimits_{k\in[K]} X_{n,k} [v_{n,k} - p_k],  \\
    r_{k} &:= \sum\nolimits_{n\in[N]}X_{n,k} p_k.
\end{align*}
The expected social welfare of $\spa(\psi)$ can be lower bounded by the total expected revenue and the total expected utility of the buyers that make purchases in the offline optimal solution, i.e.,
\begin{align*}
    \ex[\alg(\bp,I)] 
    &= \sum\nolimits_{n\in[N]} \ex [u_n] + \sum\nolimits_{k\in[K]} \ex [r_k]  \\
    &\ge \sum\nolimits_{k\in[K]} \left[\sum\nolimits_{n\in\caln_k^*}  \ex[u_n] +  \ex [r_k]\right] , 
\end{align*}
where $\caln_k^* \subseteq [N]$ is the set of buyers who decide to purchase item $k$ in the offline optimal solution. The offline social welfare can be denoted by $\opt(I) = \sum_{k\in[K]}\sum\nolimits_{n\in\caln_k^*} v_{n,k}$.
Thus, to prove Theorem~\ref{thm:oap-ub}, it is sufficient to show that
\begin{align}\label{eq:omkp-crit}
\sum\nolimits_{n\in\caln_k^*}  \ex[u_n] +  \ex [r_k] \ge \frac{1}{\alpha_k} \sum\nolimits_{n\in\caln_k^*}  v_{n,k}, \forall k\in[K].
\end{align}

For each instance $I$, let $I^{(n,k)}$ denote a modified instance, in which the valuation $v_{n,k}$ of buyer $n$ for item $k$ is set to $0$, forcing buyer $n$ not to purchase item $k$.
Given $K$ posted prices $\bp$, let $u^{(n,k)}_n$ denote the utility of buyer $n$ from \spa under the modified instance $I^{(n,k)}$. We call $u^{(n,k)}_n$ a shadow utility. 
Note that the shadow utility captures the competition from other items when buyer $n$ considers purchasing item $k$. The larger $u^{(n,k)}_n$ is, the more likely buyer $n$ is to purchase items other than $k$. Then we define the effective valuation of buyer $n$ on item $k$ as
\begin{align}
 \hat{v}_{n,k} := v_{n,k} - \min\{u^{(n,k)}_n, v_{n,k}\}.   
\end{align}
Since $u^{(n,k)}_n \ge 0$, we have $\hat{v}_{n,k} \in [0, v_{n,k}]$, which can be smaller than the lower bound $L_k$.
The shadow utility $u^{(n,k)}_n$ and effective valuation $\hat{v}_{n,k}$ satisfy the following properties. 

(i) The utility of buyer $n$ is not less than its shadow utility, i.e., $u_n \ge u^{(n,k)}_n$. If item $k$ has been sold out upon the arrival of buyer $n$, then $u_n = u^{(n,k)}_n$. Otherwise, before the arrival of buyer $n$, \spa is exactly the same under both instance $I$ and $I^{(n,k)}$. Under the instance $I$, buyer $n$ has an additional item $k$ to choose from, and thus her utility must satisfy $u_n \ge u^{(n,k)}_n$.

(ii) Buyer $n$ purchases item $k$ if her effective valuation exceeds the posted price $p_k$, i.e., $\hat{v}_{n,k} > p_k$, and the item $k$ has not been sold out. This is because $\hat{v}_{n,k} > p_k$ implies $v_{n,k} - p_k > {u}^{(n,k)}_n$, indicating that purchasing item $k$ yields a greater utility than all other items.


Recall $\caln_k^*$ contains buyers that purchase item $k$ in the offline algorithm. 
Let $(\hat{v}_{1,k}^*,\hat{v}_{2,k}^*,\dots, \hat{v}_{N_k^*,k}^*)$ denote a sequence of non-decreasing effective valuations of buyers in $\caln_k^*$, where $N_k^* = |\caln_k^*|$.
Furthermore, define a non-decreasing sequence of thresholds $(y_{0,k}, y_{1,k}, y_{2,k},\dots,y_{N_k^*,k})$ such that $y_{0,k} = 0$ and $\psi_k(y_{i,k}) := \hat{v}_{i,k}^*, \forall i\in[N_k^*]$.
Let $\rho_k(y):= \sum\nolimits_{n\in[N]} X_{n,k}$ denote the total number of sold copies of item $k$ by \spa when the static price of item $k$ is given by $\psi_k(y)$. 
Then we have 
\begin{align}\label{eq:rho-oap}
    \rho_k(y) = \sum\nolimits_{n\in[N]}X_{n,k} \ge \sum\nolimits_{i\in [N_k^*]} \mathbb{I}\{y \le y_{i,k}\}, \forall y\in[0,1].
\end{align}
To see the above inequality, if item $k$ is not sold out after the execution of the entire instance, then all buyers with effective valuations larger than $\psi_k(y)$ purchase item $k$, and thus the inequality holds. If item $k$ is sold out, then $\rho_k(y) = C_k \ge \sum\nolimits_{i\in [N_k^*]} \mathbb{I}\{y \le y_{i,k}\}$.

Conditioned on shadow utilities $\{{u}_i^{(i,k)}\}_{i\in[N_k^*]}$ from buyers in $\caln_k^*$, the expected revenue earned from item $k$ can be lower bounded by
\begin{subequations}
\label{eq:crit-oap}
\begin{align}
\ex\left[r_k |\{{u}_i^{(i,k)}\}_{i\in[N_k^*]}\right] &\ge \int\nolimits_{0}^1  \rho_k(y) \psi_k(y) dy\\
\label{eq:ineq-crit1}
& = \sum\nolimits_{i=1}^{N_k^*} \int_{0}^{y_{i,k}} \psi_k(y) dy \\
\label{eq:ineq-crit4}
&\ge \sum\nolimits_{i=1}^{N_k^*} \left[\psi_k(y_{i,k}) - (1-\frac{1}{\alpha_k}) v_{i,k}  \right]\\
&\ge \frac{1}{\alpha_k}\sum\nolimits_{i=1}^{N_k^*} v_{i,k} - \sum\nolimits_{i=1}^{N_k^*} u_i^{(i,k)},
\end{align}    
\end{subequations}
where inequality~\eqref{eq:ineq-crit1} is obtained by substituting equation~\eqref{eq:rho-oap} and exchanging the summations.
To see the inequality~\eqref{eq:ineq-crit4}, note that the designed inverse CDF function $\psi_k$ in equation~\eqref{eq:thred-oap} can ensure the following inequalities:
\begin{enumerate}
    \item when $y_{i,k} \in [0,\omega_k)$, we have 
\begin{align*}
    \psi_k(y_{i,k}) - \int_{0}^{y_{i,k}}\psi_k(x)dx \le (1 - \frac{1}{\alpha_k})L_k \le (1 - \frac{1}{\alpha_k}) v_{i,k},
\end{align*}
where $v_{i,k} \ge  L_k$ since buyer $i$ is from set $\caln_k^*$ and thus interested in item $k$. 

    \item when $y_{i,k} \in [\omega_k,1]$, we have 
\begin{align*}
    \psi_k(y_{i,k}) \le \int_{0}^{y_{i,k}}\psi_k(x)dx + (1 - \frac{1}{\alpha_k})\psi_k(y_{i,k}) \le \int_{0}^{y_{i,k}}\psi_k(x)dx + (1 - \frac{1}{\alpha_k})v_{i,k},
\end{align*}
where the last inequality holds since $v_{i,k} \ge \hat{v}_{i,k} = \psi_k(y_{i,k})$.
\end{enumerate}



Based on inequality~\eqref{eq:crit-oap}, we can have
\begin{align*}
    \ex[r_k] = \ex[\ex[r_k |\{{u}_i^{(i,k)}\}_{i\in[N_k^*]}]] &\ge \frac{1}{\alpha_k}\sum\nolimits_{i=1}^{N_k^*} v_{i,k} - \sum\nolimits_{i=1}^{N_k^*} \ex[{u}_i^{(i,k)}]\\
    &\ge \frac{1}{\alpha_k}\sum\nolimits_{i=1}^{N_k^*} v_{i,k} - \sum\nolimits_{i=1}^{N_k^*} \ex[{u}_i],
\end{align*}
which completes the proof of inequality~\eqref{eq:omkp-crit}.    
\hfill$\square$\smallskip

\subsection{Proof of Theorem~\ref{thm:lb-oap}}
\label{app:oap-lb}
We consider a setup for \ogap with $K$ items, each item having $C$ identical copies. Let $\pi$ be a permutation on the set $\{1,2,\dots,K\}$.  Let $\cala_j(C,v)$ denote a batch of $C$ identical buyers with the same valuation $v$ ($v\in[L,U]$) and interest in items from $\pi(j)$ to $\pi(K)$. We construct the hard instances as follows.

\textit{Stage I.} 
 The instance in this stage is the classic upper-triangle instance from the online matching literature~\cite{devanur2012online}. 
In particular, we consider an instance consisting of $K$ batches of buyers in the form of $\cali_L:= \cala_1(C,L)\oplus \cala_2(C,L) \oplus\dots \oplus \cala_K(C,L)$. The $(j+1)$-th batch of buyers is interested in the same items that the $j$-th batch is interested in except for the item $\pi(j)$. We claim that the  optimal
randomized online algorithm, denoted by $ \alg $, is a balancing algorithm that equally assigns buyers in each batch to their
interested items in expectation.
Specifically, the balancing algorithm $ \alg $ in expectation assigns $q_{k,j}$ buyers of the batch $\cala_j(C,L)$ to item $k$, where
\begin{align*}
    q_{k,j} = 
    \begin{cases}
      \frac{C}{K - j +1} & \pi(k) \ge j,\\
      0 & \pi(k) < j.
    \end{cases}
\end{align*}
{In addition to balancing its inventory, the optimal randomized online algorithm $ \alg $ also needs to consider a protection threshold of $\omega C$  for accepting buyers of valuation $L$ such that once the algorithm allocates one item up to the protection threshold, it would stop allocating from that item. Thus, the final utilization of item $k$ is $\min\{\omega C, \sum_{j\in[K]}q_{k,j}\}$ under $\cali_L$. The value of $\omega$ will be determined after considering instances in Stage II in such a way that the algorithm achieves the best competitive ratio. 
}

{
The online balancing algorithm $ \alg $ is optimal for the instances in Stage I. This is because any other algorithm with an imbalanced assignment of buyers to items in each batch will perform worse than the online balancing algorithm $ \alg $ under some permutation $\pi$. For instance, $\pi$ can be chosen by the adversary in such a way that the item with the lowest utilization level is discarded in the following batch.  Since the online balancing algorithm is indifferent to the permutation $\pi$, we consider w.l.o.g. that $\pi$ is an identity permutation hereinafter.
}

In contrast to the online balancing algorithm $ \alg $, the offline algorithm assigns all buyers in batch $\cala_j(C,L)$ to item $j$, and can attain $\opt(\cali_L) = L C K$. Thus, an online balancing algorithm $ \alg $ with threshold $\omega$ can achieve
\begin{align}
    \alg(\omega, \cali_L) &= L\sum\nolimits_{k=1}^K \min\left\{\omega C, \sum\nolimits_{j=1}^k \frac{ C}{K - j + 1}\right\}\nonumber\\
    &= L\sum\nolimits_{k=1}^{k_\omega}\sum\nolimits_{j=1}^k \frac{C}{K - j + 1} + L\omega C (K - k_\omega)\nonumber\\
    &\approx L C k_\omega = LC K(1-e^{-\omega}), 
\end{align}
where $k_\omega$ is the index such that $\sum_{j=1}^{k_\omega} \frac{C}{K - j + 1} = \omega C$. Thus $\omega \approx \ln\frac{K}{K-k_\omega}$, and $k_\omega \approx K (1-e^{-\omega})$ as $K \to \infty$. The last equation is obtained by observing that $\sum_{k=1}^{k_\omega}\sum_{j=1}^k \frac{1}{K - j + 1} = \sum_{j=1}^{k_\omega}\sum_{k=j}^{k_\omega} \frac{1}{K - j + 1} = k_\omega - (K- k_\omega)\sum_{j=1}^{k_\omega}\frac{1}{K - j + 1} = k_\omega - (K- k_\omega) \omega$.



\textit{Stage II.} 
At the end of Stage I, the items from $K(1-e^{\omega})$ to $K$ have sold $\omega C$ units of items. 
The arrivals of Stage II are constructed similarly to the worst-case instance of the online selection problem in the proof of Lemma~\ref{lem:osp-lb}, and only interested in one of the last $K e^{\omega}$ items. 
Specifically, an instance of Stage II with maximum value $v$ ($v\in(L,U]$) is denoted by $\cali_v := \cali_L \oplus \cala_{K}(C, L+\epsilon)\oplus \cala_{L}(C, L+2\epsilon)\oplus\dots\oplus \cala_{K}(C, v)$, which consists of a sequence of batches that are only interested in the item $K$ and their valuations continuously increase from $L$ to $v$. 
Then the offline algorithm under instance $\cali_v$ assigns batch $\cala_j(C,L)$ to item $j$ for $j\in[K-1]$ and assigns the batch $\cala_K(C,v)$ to item $K$, achieving the optimal social welfare $\opt(\cali_v) = L C (K - 1) + v C.$
Let $g(v)$ denote the average representative function of the item $K$ by running a randomized online algorithm under instance $\cali_v$, where $g(L):= \omega C$ is the maximum sold units in Stage I. 
The expected social welfare of the online algorithm characterized by $g$ can be described as $\alg(g, \cali_v) = LC k_\omega + \int_{L}^v u d g(u),$
where the first term is return from the instance in Stage I and the second term is return in Stage II.
Any $\alpha$-competitive online algorithm must satisfy $\alg(g,\cali_v) \ge \frac{1}{\alpha} \opt(\cali_v), \forall v\in[L,U]$. This gives
\begin{subequations}
 \begin{align}
    \label{eq:lb-omk1}
    L C k_\omega &\ge \frac{1}{\alpha}L C K,\\
     \label{eq:lb-omk2}
    LC k_\omega + \int_{L}^v u d g(u) &\ge \frac{1}{\alpha} \left[L C (K - 1) + v C \right], \quad \forall v\in(L,U].  
\end{align}   
\end{subequations}
Equation~\eqref{eq:lb-omk1} gives us a lower bound of the competitive ratio $\alpha \ge \frac{e^\omega}{e^\omega - 1}$, and equation~\eqref{eq:lb-omk2} gives
\begin{align*}
     g(v) \ge  \frac{C}{\alpha} + \frac{1}{v} \int_{L}^v g(u) du + \frac{L \omega C}{v} - \frac{LC}{v\alpha} \ge \omega C + \frac{C}{\alpha}\ln \frac{v}{L}, \quad \forall v\in(L,U]. 
\end{align*}
Since $g(U) \le C$, we have $\alpha \ge \frac{\ln\theta}{1 -\omega}$.
Thus, the lower bound of \ogap is
\begin{align*}
    \alpha \ge \min_{\omega \in [0,1]} \max\left\{\frac{e^\omega}{e^\omega - 1}, \frac{\ln\theta}{1 -\omega} \right\},
\end{align*}
which is achieved when the threshold $\omega$ is the solution of $\frac{e^\omega}{e^\omega - 1} = \frac{\ln\theta}{1 -\omega}$. 
\hfill$\square$\smallskip

\section{Proofs for Online Selection with Convex Cost (\oscc)}

\subsection{Proof of Theorem~\ref{thm:oscc-ub}}
\label{app:oscc-ub}
$\spa(\psi)$ samples a price $p$ from a distribution $\psi$, and posts this price to all buyers. 
Each buyer $n$ has a private valuation $v_n$ on the item, and decides to buy the item if $v_n \ge p$ and the number of sold items has not exceeded the effective capacity. 
Let $X_n\in\{0,1\}$ denote the purchase decision of buyer $n$. Then buyer $n$ can obtain a utility $u_n = (v_n - p)X_n$ and the seller can collect a total profit of $r = \sum_{n\in[N]} p X_n - f(\sum_{n\in[N]} X_n)$.
Given an instance $I=\{v_1,\dots,v_N\}$, the social welfare attained by $\spa(\psi)$ under instance $I$ can be described by
\begin{align*}
\mathbb{E}[\alg(p,I)] &= \ex\left[\sum\nolimits_{n\in[N]} X_n (v_n - p) + \sum\nolimits_{n\in[N]} p X_n - f(\sum\nolimits_{n\in[N]} X_n)\right] := \ex \left[\sum\nolimits_{n\in[N]} u_n + r \right].
\end{align*}
Let $(v_1^*, v_2^*, \dots, v_{N^*}^*)$ denote the sequence of buyers' valuations selected by the offline algorithm and arranged in a non-decreasing order. These are the $N^*$-maximum valuations from all buyers in $I$. 
Furthermore, define a sequence of non-decreasing thresholds $(x_0, x_1,\dots, x_{N^*})$ such that $x_0 = 0$ and $\psi(x_i) = v_i^*, i\in[N^*]$.
Let $\rho(x):= \sum\nolimits_{n\in[N]} X_n$ denote the total number of sold items by \spa when the static price is given by $\psi(x)$. We can have 
\begin{align}\label{eq:oscc-rho}
    \rho(x) = \sum\nolimits_{n\in[N]}X_{n} \ge \sum\nolimits_{i\in [N^*]} \mathbb{I}\{x \le x_i\}, \forall x\in[0,1].
\end{align}

Next, we can lower bound the expected social welfare of $\spa(\psi)$. Note that $v_1^*$ is the minimum valuation selected by the offline algorithm. Therefore, the total number of sold items $N^*$ by the offline algorithm cannot exceed $\yinv(v_1^*)$. We consider the following two cases.
\paragraph{Small demand case: $N^* < \yinv(v_1^*)$.} In this case, there is a total of $N^*$ buyers in the instance since if there were an additional buyer with arbitrary valuation within $[L,U]$, the offline algorithm will select it.  
Thus, offline algorithm is to select all buyers, while the \spa selects all buyers (by setting $p = L$) with probability at least $\frac{1}{\alpha}$ based on the probability density function~\eqref{eq:oscc-dist}. Thus,
\begin{align}
    \ex[\alg(p,I)] \ge \frac{1}{\alpha} \cdot \alg(L,I) = \frac{1}{\alpha}\cdot\opt(I).  
\end{align}

\paragraph{Large demand case: $N^* = \yinv(v_1^*)$.} In this case, the offline algorithm is to exactly select $\yinv(v_1^*)$ buyers, and the optimal social welfare is $\opt(I) = \sum_{i\in [N^*]} v_i^* - f(N^*)$. Then we can lower bound the expected social welfare by
\begin{subequations}
\begin{align}
\label{eq:oscc-ineq0}
\mathbb{E}[\alg(p,I)] &\ge  \int_{0}^{x_1} h(\psi(x)) dx +\sum\nolimits_{j=2}^{N^*} \int_{x_{j-1}}^{x_j} [\rho(x) \psi(x)  - f(\rho(x))]dx\\
\label{eq:oscc-ineq1}
&\ge \int_{0}^{x_1} h(\psi(x)) dx  + \sum\nolimits_{i=2}^{N^*} \int_{x_{1}}^{x_{i}} \frac{h(\psi(x))}{\yinv(\psi(x))} dx,\\
\label{eq:oscc-ineq2}
&\ge \frac{N^* \psi(x_1) - f(N^*)}{\alpha} + \sum\nolimits_{i=2}^{N^*} \frac{\psi(x_i) - \psi(x_1)}{\alpha} \\
&= \frac{1}{\alpha} \left[\sum\nolimits_{i\in[N^*]} v_i^* - f(N^*)\right] = \frac{1}{\alpha} \opt(I).
\end{align}    
\end{subequations}
To see the first part of inequality~\eqref{eq:oscc-ineq0}, we note that $\rho(x) \ge  N^*$ when $x < x_1$. With static price $\psi(x)$, $N^*$ selected buyers can at least achieve $h(\psi(x)) = \yinv(\psi(x)) \psi(x) - f(\yinv(\psi(x)))$ when the first $\yinv(\psi(x))$ buyers are with valuation $\psi(x)$ and the remaining $N^* - \yinv(\psi(x))$ buyers' the valuations are exactly the same as the marginal production cost. The second part of inequality~\eqref{eq:oscc-ineq0} is because valuations of all selected buyers are greater than $\psi(x)$.
The inequality~\eqref{eq:oscc-ineq1} holds since
\begin{subequations}
\begin{align}
\sum\nolimits_{j=2}^{N^*} \int_{x_{j-1}}^{x_j} [\rho(x) \psi(x)  - f(\rho(x))]dx & \ge  \sum\nolimits_{j=2}^{N^*}\int_{x_{j-1}}^{x_{j}}  \rho(x)[\psi(x) - \frac{f(N^*)}{N^*}]dx \\
&\ge \sum\nolimits_{i=2}^{N^*} \int_{x_{1}}^{x_{i}}  [\psi(x) - \frac{f(N^*)}{N^*}]dx  \\
&\ge \sum\nolimits_{i=2}^{N^*} \int_{x_{1}}^{x_{i}} \frac{h(\psi(x))}{\yinv(\psi(x))} dx,
\end{align}    
\end{subequations}
where the first inequality is due to the convexity of the production function $\frac{f(N^*)}{N^*} \ge \frac{f(\rho(x))}{\rho(x)}$ when $\rho(x) \le N^*$, and second equality is obtained by substituting the inequality~\eqref{eq:oscc-rho} and exchanging the summation, and the last inequality is by observing that  $\frac{f(N^*)}{N^*} \le \frac{f(\yinv(\psi(x)))}{\yinv(\psi(x))}$, where $\yinv(\psi(x)) \ge N^*$ when $x \ge x_1$. 

Finally, the inequality~\eqref{eq:oscc-ineq2} is based on the design of the inverse CDF $\psi$. 
In particular, 
\begin{subequations}
\begin{align}
    \int_0^{x_1} h(\psi(x))dx &\ge \frac{h(\psi(x_1))}{\alpha} = \frac{N^* \psi(x_1) - f(N^*)}{\alpha},\\
    \int_{x_{1}}^{x_{i}} \frac{h(\psi(x))}{\yinv(\psi(x))} dx&\ge 
    \frac{1}{\alpha}\int_{x_{1}}^{x_{i}} \psi'(x) dx = \frac{\psi(z_i) - \psi(z_1)}{\alpha}, \forall i = 2,\dots,C.
\end{align}    
\end{subequations}
This completes the proof.

\subsection{Proof of Theorem~\ref{thm:oscc-lb}}
\label{app:oscc-lb}
Consider a family of continuously increasing instances $\{I_v\}_{v\in[L,U]}$, which has been constructed in Section~\ref{sec:osp-lb} to prove the lower bound of the online selection problem.
Recall $I_v$ consists of a sequence of buyers whose valuations continuously increase from $L$ to $v$, and the total number of buyers with the same valuation is $C$.
Let $G(v):[L,U]\to[0,1]$ denote the CDF of the static price, and then $G(v)$ can model all possible randomized static posted price algorithms. 

Under the instance $I_{L}$, the offline social welfare is $\opt(I_{L}) = h(L) = \max_{y\in \calc} L y - f(y)$, and the expected social welfare of the static pricing algorithm with CDF $G$ is $\ex[\alg(G, I_L)] = G(L) h(L).$
An $\alpha$-competitive algorithm must ensure $\ex[\alg(G, I_{L})] \ge \frac{1}{\alpha} \opt(I_{L})$, and thus $G$ must satisfy $G(L) \ge \frac{1}{\alpha}$.
$G(L)$ is the probability that the algorithm chooses $L$ as the static price, which is in fact a greedy algorithm that selects all buyers whose valuations are less than the marginal production cost.

Under the instance $I_v, v\in(L,U]$, the optimal social welfare is $\opt(I_{v})  = h(v)$ and the expected social welfare of a static pricing algorithm can be computed by $\ex[\alg(G, I_v)] = G(L) h(L) + \int_{L}^v h(u) d G(u)$.
To ensure $\ex[\alg(G, I_{v})] \ge \frac{1}{\alpha} \opt(I_{v})$, $G$ must satisfy $G(L) h(L) + \int_{L}^{v} h(u) d G(u) \ge \frac{1}{\alpha} h(v), \forall v\in(L,U],$
which, through integral by parts, can be equivalently transformed to $G(v) h(v)- \int_{L}^{v} G(u)h'(u)du \ge \frac{1}{\alpha} h(v), \quad \forall v\in(L,U]$. 
Combining the above equation and $G(L) \ge 1/\alpha$, we claim that if there exists an $\alpha$-competitive static pricing algorithm, then there must exist $G$ such that 
\begin{align}\label{eq:lb-func}
    G(v) h(v) - \int_{L}^{v} G(u)h'(u)du \ge \frac{1}{\alpha} h(v),  \quad \forall v\in[L,U].
\end{align}

Based on Gronwall's inequality, we can equivalently have, $\forall v\in[L,U]$,
\begin{subequations}
\begin{align}
    G(v) &\ge \frac{1}{\alpha} + \frac{1}{h(v)}\int_{L}^{v} G(u)h'(u)du\\
    &\ge \frac{1}{\alpha} + \frac{1}{\alpha h(v)}\int_{L}^{v} h'(u) \exp\left(\int_{u}^v   \frac{h'(s)}{h(s)} ds\right)du,\\
    &= \frac{1}{\alpha} + \frac{1}{\alpha}\int_{L}^{v} \frac{h'(u)}{h(u)} du,\\
    &= \frac{1}{\alpha} + \frac{1}{\alpha} \ln\frac{h(v)}{h(L)},
\end{align}    
\end{subequations}
Since the CDF satisfies $G(U) = 1$, we have $\alpha \ge 1 + \ln\frac{h(U)}{h(L)}$. This completes the proof. 

\end{document}